\documentclass[11pt,a4paper]{article}
\pdfoutput = 1
\usepackage{jheppub}
\usepackage{enumitem}
\usepackage{amsmath}
\usepackage{amsfonts}
\usepackage{graphicx}

\title{Random Walks in Rindler Spacetime and String Theory at the Tip of the Cigar}

\author[a]{Thomas G. Mertens,}
\author[a]{Henri Verschelde}
\author[b]{and Valentin I. Zakharov}
\affiliation[a]{Ghent University, Department of Physics and Astronomy\\
Krijgslaan, 281-S9, 9000 Gent, Belgium}
\affiliation[b]{ITEP, B. Cheremushkinskaya 25, Moscow, 117218 Russia,\\
Max-Planck Institut f¨ur Physik, 80805 M¨unchen, Germany,\\
Moscow Inst Phys \& Technol, Dolgoprudny, Moscow Region, 141700 Russia 
}
\emailAdd{thomas.mertens@ugent.be}
\emailAdd{henri.verschelde@ugent.be}
\emailAdd{vzakharov@itep.ru}
\abstract{In this paper, we discuss Rindler space string thermodynamics from a thermal scalar point of view as an explicit example of the results obtained in \cite{theory}. We discuss the critical behavior of the string gas and interpret this as a random walk near the black hole horizon. Combining field theory arguments with the random walk path integral picture, we realize (at genus one) the picture put forward by Susskind of a long string surrounding black hole horizons. We find that thermodynamics is dominated by a long string living at string-scale distance from the horizon whose redshifted temperature is the Rindler or Hawking temperature. We provide further evidence of the recent proposal for string theory at the tip of the cigar by comparing with the flat space orbifold approach to Rindler thermodynamics. We discuss all types of closed strings (bosonic, type II and heterotic strings).}

\begin{document}

\maketitle

\section{Introduction}

In \cite{theory}, we analyzed the method set forth by the authors of \cite{Kruczenski:2005pj} to analyze the near-Hagedorn thermodynamics of string theory directly from the string path integral. The method explicitly describes the random walk picture of high-temperature string thermodynamics: after heating up a gas of closed strings, the constituents coalesce into a single long, highly excited closed string.\footnote{Actually this depends on the compactness of the different dimensions \cite{Deo:1989bv}. For $d\geq3$ a single string dominates, for $d=0$ multiple-string configurations contribute, whereas for $d=1$ and $d=2$ a more detailed analysis is required \cite{Deo:1989bv}\cite{Deo:1988jj}.} This long string behaves as a random walker in a fixed background. From a Euclidean point of view, the long string is described by the thermal scalar \cite{Atick:1988si}. This is a complex scalar field that combines the winding $\pm 1$ stringy excitations around the Euclidean time circle. We noted that this random walk receives corrections compared to the naive `worldsheet dimensional reduction' to a particle theory. The random walk picture also emerges when considering black hole horizons \cite{Susskind:1993ws}\cite{Susskind:2005js}. The long string(s) wraps the horizon and effectively form the stretched horizon. This is a microscopic stringy candidate for the black hole membrane as it is called in the earlier literature.\footnote{See e.g. \cite{Thorne:1986iy} and references therein.} This picture and the related correspondence principle have been studied extensively in the past by numerous authors (see e.g. \cite{Halyo:1996vi}\cite{Maldacena:1996ds}\cite{Halyo:1996xe}\cite{Halyo:2001us}\cite{Halyo:2003bt}\cite{Sen:2004dp}\cite{Giveon:2006pr}\cite{BatoniAbdalla:2007zv}\cite{Sasai:2010pz}). The stretched horizon is located at a string length outside the black hole horizon. It is our goal to analyze this picture further for a string gas close to the horizon or more specifically for a string gas in Rindler space using the explicit methods developed earlier. Recently, in light of the so-called firewall-paradox \cite{Almheiri:2012rt},\footnote{See \cite{Braunstein:2009my} for an earlier account of this.} it has become increasingly important to better understand how string theory behaves near black hole horizons. 

The methods developed only consider the genus one worldsheet (torus) and are thus firmly rooted in perturbative string theory. We are aware that this limits the applications. In particular it has been argued \cite{Susskind:2005js} that close to the black hole horizon, one cannot ignore the higher genus contributions. 
Due to a lack of analytical methods to analyze non-perturbative string theory in this regime, we take the genus one results as a guide to what actually happens near black hole horizons. The genus one approach to Rindler string thermodynamics has been considered quite extensively in the past (see e.g. \cite{Dabholkar:1994ai}\cite{Lowe:1994ah}\cite{Parentani:1989gq}\cite{Emparan:1994bt}\cite{McGuigan:1994tg}).\\
The methods we will use are a mixture of two approaches: firstly we use a worldsheet Fourier series expansion of the string path integral. This provides an explicit random walk picture and it can be interpreted as the long string in the Lorentzian picture. Secondly, we use the field theory action of the thermal scalar. This field theory point of view will allow us to explicitly see the corrections to the particle action that the first approach misses. Also, off-shell questions are accessable using field theory. When we combine these two approaches, the difficulties of either approach are better understood and we obtain a realization of the (genus one) long string surrounding the black hole as was anticipated and argued by Susskind several years ago.

This paper is organized as follows.\\
In section \ref{pathderiv} we recapitulate the results from \cite{theory}. In section \ref{rindler} we discuss Rindler space thermodynamics and its link to black hole horizons. In section \ref{alphaprime} we elaborate on the higher order $\alpha'$ corrections for the specific case of Rindler space. We discuss this in general first and then we use the link to the cigar CFT as was recently proposed in \cite{Giveon:2013ica}\cite{Giveon:2012kp}\cite{Giveon:2014hfa}. Using this knowledge, we analyze the thermal scalar in Rindler space in section \ref{critical} for bosonic, type II and heterotic strings. In particular we determine the Hagedorn temperature, the location of the random walk and whether the free energy remains finite or not. We discuss the Hagedorn behavior and the density of states of the Rindler string and show that the free energy is dominated by a long string at string-scale distance from the horizon with redshifted critical temperature equal to the Rindler temperature. In section \ref{orbifold} we make the link between the critical behavior of flat space $\mathbb{C}/\mathbb{Z}_N$ orbifold models and angular orbifolds of the $SL(2,\mathbb{R})/U(1)$ model and in particular we will find that the non-standard momentum-winding duality of bosonic strings on the cigar \cite{Dijkgraaf:1991ba} has an important manifestation on the flat $\mathbb{C}/\mathbb{Z}_N$ orbifold. In section \ref{unitarity} we identify the Rindler quantum numbers with those from the cigar model. We will discuss the unitarity constraints and how they are translated to the Euclidean Rindler case. We present discussions, some remaining open problems and our conclusions in sections \ref{discussion}, \ref{open} and \ref{conclusion}. Several illustrative and technical calculations are gathered in the appendices.

\section{Set-up and plan}
\label{pathderiv}
The authors of \cite{Kruczenski:2005pj} have given an explicit path integral picture of the thermal scalar. We have extended their result in \cite{theory}. Let us recapitulate the results here.\\
Performing a naive worldsheet dimensional reduction of the torus partition function, we found that the torus path integral on the thermal manifold reduces to
\begin{equation}
\label{act0} 
Z_p = 2\int_0^\infty \frac{d \tau_2}{2\tau_2} \int \left[ \mathcal{D}X \right] \sqrt{\prod_{t} \det G_{ij}} \exp - S_p(X) 
\end{equation}
where 
\begin{equation}
\label{act}
S_p = \frac{1}{4\pi \alpha'}\left[ \beta^2 \int_0^{\tau_2} dt G_{00} + \int_0^{\tau_2} dt G_{ij} \partial_t X^i \partial_t X^j\right].
\end{equation}
The full string partition function has been reduced to a partition function for a non-relativistic particle moving in the purely spatial submanifold. The time evolution of the particle in its random walk is identified with the spatial form of the long highly excited string. We view this as a realization of the Wick rotation: the long string in real spacetime has a form described by the above random walk.\\
The free energy of a gas of strings can then be identified with the single string partition function as \cite{Polchinski:1985zf}
\begin{equation}
F = -\frac{1}{\beta} Z_p.
\end{equation}
An alternative route we followed was the field theory of the particle modes. Using this theory for the thermal scalar, we were able to see that correction terms to the above particle action were in order. The thermal scalar action is given by (to lowest order in $\alpha'$)
\begin{equation}
\label{lowestFT}
S \sim \int d^{D-1}x \sqrt{G_{ij}}\sqrt{G_{00}}e^{-2\Phi}\left(G^{ij}\partial_{i}T\partial_{j}T^{*} + \frac{R^2G_{00}}{\alpha'^2}TT^{*} + m^2TT^{*}\right),
\end{equation}
where $m^2$ is the tachyon mass$^2$ in flat space whose precise value will be given below. From this action, the one-loop free energy is given by
\begin{align}
\label{FT}
\beta F &= - \int_{0}^{+\infty}\frac{dT}{T} \text{Tr} e^{-T\left(-\nabla^{2}+m_{local}^2 - G^{ij}\frac{\partial_{j}\sqrt{G_{00}}}{\sqrt{G_{00}}}\partial_{i}\right)} \\
\label{randwalk}
&= - \int_{0}^{+\infty}\frac{dT}{T}\int_{S^{1}} \left[\mathcal{D}x\right]\sqrt{G}e^{-\frac{1}{4\pi\alpha'}\int_{0}^{T}dt\left(G_{ij}(x)\dot{x}^{i}\dot{x}^{j} + 4\pi^2\alpha'^2\left(m_{local}^2 + K(x\right)\right)}.
\end{align}
We denote the operator in brackets in the exponential in the first equation as $\hat{\mathcal{O}}$ in what follows. \\
We have also collected the `local' mass terms in
\begin{align}
\label{mlocal}
m_{local}^2 &= -\frac{4}{\alpha'} + \frac{R^2G_{00}}{\alpha'^2}, \quad \text{for bosonic strings}, \\
m_{local}^2 &= -\frac{2}{\alpha'} + \frac{R^2G_{00}}{\alpha'^2}, \quad \text{for type II superstrings}, \\
m_{local}^2 &= -\frac{3}{\alpha'} +\frac{1}{4R^2G_{00}}+\frac{R^2G_{00}}{\alpha'^2}, \quad \text{for heterotic strings}.
\end{align}
The function $K(x)$ denotes the following metric combination\footnote{$\nabla^2$ is the Laplacian on the spatial submanifold.}
\begin{equation}
\label{K}
K(x) =-\frac{3}{16}\frac{G^{ij}\partial_iG_{00}\partial_jG_{00}}{G_{00}^2} + \frac{\nabla^2 G_{00}}{4G_{00}}
\end{equation}
and this represents the effect of removing the $\sqrt{G_{00}}$ from the measure in the field theory action. Going from (\ref{FT}) to (\ref{randwalk}) requires some delicate manipulations that we discussed in \cite{theory}. Equation (\ref{randwalk}) can then be identified with (\ref{act0}) and (\ref{act}) and hence we can see which correction terms are needed in the particle action. The correction terms are of three different types:
\begin{itemize}
\item{Firstly we have a correction term coming from the mass of the closed string tachyon and this is of the following form
\begin{equation}
\Delta S = -\frac{\beta_{H}^2\tau_2}{4\pi\alpha'}.
\end{equation}
For bosonic strings $\beta_{H}^2 = 16\pi^2\alpha'$, for type II superstrings $\beta_{H}^2 = 8\pi^2\alpha'$ and for heterotic strings $\beta_{H}^2 = 12\pi^2\alpha'$. 
For bosonic and type II strings, this is the flat space Hagedorn temperature but for heterotic strings this is not the case. By abuse of notation, we will nonetheless denote this term with $\beta_H^2$.
}
\item{Secondly we have a correction coming from the $G_{00}$ component as explained in \cite{theory}: 
\begin{equation}
\label{corr}
\Delta S = \frac{1}{4\pi\alpha'}\int_{0}^{\tau_2}dt 4\pi^2\alpha'^2 K(x).
\end{equation}
}
\item{Finally we could have order-by-order $\alpha'$ correction terms of the field theory action. These are of course not present in (\ref{randwalk}) and it is difficult to say anything specific about these in the general case. Nonetheless we will obtain these corrections for the specific case of Euclidean Rindler space.}
\end{itemize}
In \cite{theory}, we only analyzed flat backgrounds explicitly. For a general curved background however, we have much less control on what precisely happens. Two questions need to be answered:
\begin{itemize}
\item{Is there a winding mode that becomes tachyonic when exceeding a specific temperature? We know this to be the case for flat backgrounds, but does it hold for more general (either topologically or geometrically non-trivial) backgrounds? Can we establish a regime where this thermal scalar dominates the thermodynamical quantities?}
\item{Can we get a handle on what other $\alpha'$ correction terms need to be added to the thermal scalar action for a general background?}
\end{itemize}
We will find answers to both of these questions in our study of Rindler space. While this space is geometrically quite easy, the description of string thermodynamics in terms of the Rindler observer is not straightforward. Nevertheless we will be able to explicitly solve the critical behavior of the Rindler string.
We will make contact with some previous results regarding string thermodynamics in Rindler spacetime and we will obtain a prediction of the Rindler Hagedorn temperature for all types of (conventional) string theories.

\section{Rindler thermodynamics}
\label{rindler}
We will be interested in computing the one-loop free energy of strings in the critical regime. In general one knows that a black hole is surrounded by a thin membrane called the stretched horizon. This Planck (or string) sized object is outside the reach of quantum field theory in curved spacetimes. This can be seen by e.g. the UV divergence of one-loop thermodynamical quantities \cite{'tHooft:1984re}\cite{Barbon:1994ej}\cite{Emparan:1994qa} of a gas of particles in the black hole geometry. When looking further from the horizon, a thermal zone is found that is stretched over roughly one Schwarzschild radius radially outwards. This is also the region where we can approximate the black hole background by Rindler spacetime as we now demonstrate \cite{Susskind:2005js}. Consider a D-dimensional Schwarzschild spacetime\footnote{We take the Schwarzschild black hole as an example to demonstrate our point in what follows, even though it is not an exact string background. In general, most uncharged black holes have a Rindler near-horizon limit. Rindler space on the other hand is an exact string background.}
\begin{equation}
ds^2 = -\left(1-\frac{2GM}{r}\right)dt^2 + \left(1-\frac{2GM}{r}\right)^{-1}dr^2 + d\mathbf{x}^2_{\perp}.
\end{equation}
If we define a new coordinate $\rho = \sqrt{8GM(r-2GM)}$ and focus on the near horizon geometry, we reduce this metric to
\begin{equation}
ds^2 = -\left(\frac{\rho^2}{(4GM)^2}\right)dt^2 + d\rho^2 + d\mathbf{x}^2_{\perp}.
\end{equation}
This is the Rindler metric. The above derivation holds as long as
\begin{equation}
\rho \ll 4GM.
\end{equation}
Our goal is to analyze the near-horizon behavior of the one-loop free energy in string theory. Note that to describe thermodynamics, we need to choose a preferred time coordinate. In this case it is Rindler time (or Schwarzschild time), so we describe thermodynamics as perceived by fiducial observers in the spacetime. In particular, space ends at the horizon by a smooth capping of the cigar. In string theory, one does not expect a UV divergence in the free energy, but we should be careful about IR divergences which can (and will) occur. Considering Rindler spacetime as a string background, all background fields are turned off except the metric whose Euclidean section takes the following form
\begin{equation}
\label{Rindmetric}
ds^2 = \frac{\rho^2}{(4GM)^2}d\tau^2 + d\rho^2 + d\mathbf{x}^2_{\perp}.
\end{equation}
To avoid a conical singularity at $\rho=0$, the Euclidean time coordinate needs to have the identification $\tau \sim \tau + \beta_R$ where 
$\beta_R = 8\pi GM$. We will refer to this temperature as the canonical Rindler temperature. Euclidean Rindler space is then the same as flat space in polar coordinates. This temperature coincides with the Hawking temperature of the original black hole since we did not change the temporal coordinate. \\

We want to remark that this temperature is a global property of the space. The local temperature is equal to $\beta_{R,local} = \beta_{R}\sqrt{G_{00,Rindler}}$. This is the temperature as measured by local observers. In particular, the canonical Rindler temperature itself is measured by an observer located at $G_{00} = 1$. In this case this is at $\rho = 4GM$. \\

Since our intention is to study thermodynamics at the canonical Rindler temperature, we are actually looking at the stringy Unruh effect and hence the vacuum we are considering is the Minkowski vacuum. As is well-known, the description of this vacuum by a fiducial observer corresponds to a thermal state at the canonical Rindler temperature. \\

Calculating thermodynamical properties in such spaces presents some complications.
In field theory, varying $\beta$ results in conical spaces. Although one would at first sight think that a conical space represents non-equilibrium,\footnote{And thus one cannot infinitesimally vary the temperature to nearby equilibrium configurations, as one is instructed to do in thermodynamical calculations.} it can be proven that if one \emph{only} varies the temperature and hence fixes the transverse horizon area \cite{Susskind:1994sm}\cite{Carlip:1993sa}, one indeed gets these conical spaces. So it is consistent for equilibrium phenomena to infinitesimally vary $\beta$ as long as one sets it equal to $8\pi GM$ in the end. \\
In string theory the situation is more problematic: besides the previous complication, one also has the problem that for $\beta \neq \frac{8\pi GM}{N}$ with $N \in \mathbb{N}$, the worldsheet theory is not conformal and thus inconsistent.\footnote{Corresponding to the fact that string theory in its standard formulation cannot be taken off-shell.} This was in fact the rationale behind the orbifold approximations to Rindler spacetime thermodynamics by the authors of \cite{Dabholkar:1994ai}\cite{Lowe:1994ah}. We will come back to this further on. In our case, we leave $\beta$ general, and in spite of the fact that the starting point of the derivation in section \ref{pathderiv} is only valid for one value of $\beta$, we will interpret our final expression as an off-shell continuation of the conformal result.\footnote{Although such an interpretation is not really necessary for much of what is to follow.} Moreover, we have seen that one can also get the particle path integral from the field theory action and this action obviously \emph{can} be off-shell. We will return to this topic further on. \\

The free energy is a \emph{globally} defined concept and we do not make any assertions about the local properties of thermodynamic quantities. As discussed by \cite{Emparan:1994bt}, locality in string theory is a delicate topic and also even in field theory the thermal properties in black hole backgrounds are all global properties.\footnote{See \cite{Calcagni:2013eua} for a very recent account on string nonlocality.} Local arguments can be a good guide in a general background, but in a cigar-shaped background they are misleading: local reasoning predicts a divergence simply because the thermal circle keeps on shrinking below the Hagedorn radius. We will see that this is wrong in general and that the thermal quantities for most string types are tachyon-free. We want to notice that this dramatic failure of local reasoning is typical for cigar-shaped backgrounds. For manifolds with topologically stable thermal circles on the other hand (like Euclidean $AdS_3$), local reasoning is qualitatively good (also see \cite{Rangamani:2007fz}).

\section{Corrections in $\alpha'$ to the thermal scalar action}
\label{alphaprime}
One of our main interests lies in getting a handle on possible higher $\alpha'$ corrections to the thermal scalar action. 
We will first discuss this in general using only general covariance as a guide. Further on we will make the link with the cigar $SL(2,\mathbb{R})/U(1)$ CFT model where we will precisely pinpoint what these corrections look like.

\subsection{General analysis}
\label{general}
\begin{figure}[h]
\centering
\includegraphics[width=11cm]{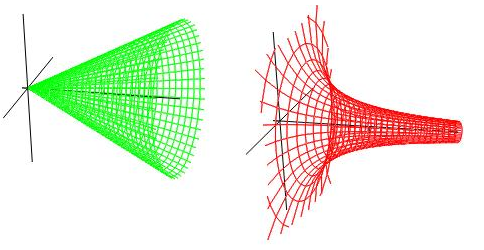}
\caption{Left figure: size of the thermal circle in Rindler space as a function of radial distance increasing towards the right. Right figure: size of the thermal circle in the T-dual of Rindler space as a function of radial distance increasing towards the right.}
\label{thermalcircles}
\end{figure}

We remind the reader that the thermal scalar action is given by the effective action for the first discrete momentum mode in the T-dual background, i.e. $n=\pm1$ and $w=0$. Corrections to the thermal scalar action can thus be analyzed by considering the discrete momentum action of the T-dual background. 
In this section we analyze the general form of possible $\alpha'$ corrections to the thermal scalar action in Rindler space and we discuss whether we can identify a regime where these, if they are present, are subdominant.
Considering Euclidean Rindler space (\ref{Rindmetric}), the T-dual Ricci tensor has components
\begin{equation}
\tilde{R}^{00} = -\frac{2}{(4GM)^2}, \quad  \tilde{R}^{\rho\rho} = -\frac{2}{\rho^2}, \quad \tilde{R}^{0\rho}=0
\end{equation}
and the T-dual Ricci scalar $\tilde{R} = -\frac{4}{\rho^2}$. Note that there is a curvature singularity at $\rho=0$ (for the string metric). The T-dual dilaton field is given by
\begin{equation}
\partial_\rho \tilde{\Phi} = -\frac{1}{\rho}.
\end{equation}
Some possible terms that could appear in the field theory action are given by
\begin{align}
m^2 TT^{*} &= -\frac{4}{\alpha'}TT^{*}, \quad \text{bosonic}\quad \text{or} \quad m^2 TT^{*} = -\frac{2}{\alpha'}TT^{*}, \quad \text{type II},\\
\tilde{G}^{\mu\nu}\partial_{\mu}T\partial_{\nu}T^{*} &=  \frac{w^2\beta^2\rho^2}{4\pi^2(4GM)^2\alpha'^2}TT^{*} + \partial_\rho T \partial_\rho T^{*} = \frac{\rho^2}{\alpha'^2}TT^{*} + \partial_\rho T \partial_\rho T^{*}, \\
\label{curv0}
\tilde{R} TT^{*} &= -\frac{4}{\rho^2}TT^{*}, \\
\label{masscorr}
\alpha'\tilde{R}^{\mu\nu}\partial_\mu T \partial_{\nu} T^{*} &= -\frac{w^2\beta^2}{2\pi^2(4GM)^2\alpha'}TT^{*} - 2\frac{\alpha'}{\rho^2}\partial_\rho T \partial_\rho T^{*} = -\frac{2}{\alpha'}TT^{*} - 2\frac{\alpha'}{\rho^2} \partial_\rho T \partial_\rho T^{*}
\end{align}
and
\begin{align}
\label{curv1}
\partial_\mu \tilde{\Phi} \partial^{\mu} \tilde{\Phi} TT^{*} &= \frac{1}{\rho^2}TT^{*}, \\
\label{curv2}
\alpha'\partial^{\mu}\tilde{\Phi} \partial^{\nu} \tilde{\Phi} \partial_{\mu}T\partial_{\nu}T^{*} &= \frac{\alpha'}{\rho^2}\partial_\rho T \partial_\rho T^{*},
\end{align}
since $\partial_0 = i\frac{\beta}{2\pi\alpha'} = i (4GM) / \alpha'$ and $\left|w\right|=1$. Note that if we do not fix the temperature to the Hawking temperature, we get $\beta^2$ corrections to the action but not other powers of $\beta$. This will be relevant further on. We see that higher order $\alpha'$ terms are not suppressed due to two reasons:
\begin{itemize}
\item[1.] The T-dual geometry has curvature that blows up at $\rho = 0$. Comparing the original Rindler space with its T-dual, the coordinate singularity of the original black hole gets transformed into a curvature singularity. If we would be interested in the far region from the horizon, we could still neglect these terms. So only for $\rho$ large, will these terms be subdominant. Some terms of this type are (\ref{curv0}), (\ref{curv1}), (\ref{curv2}) and the second term of (\ref{masscorr}).
\item[2.] Secondly, the temperature is not string scale, but is equal to the Hawking temperature. This provides extra contributions (like the first term in (\ref{masscorr})) in the action that are of the same order in $\alpha'$ as the naive lowest $\alpha'$ action whose general form was given in (\ref{lowestFT}).
\end{itemize}
This conclusion holds in general: the $\alpha'$ corrections are subdominant if the T-dual curvature radius is (much) larger than the string length and the temperature is string scale. So if the thermal circle does not deviate much from flat space, its T-dual space will also be quite well-behaved and the $\alpha'$ corrections are subdominant. An example is $AdS$ space as we will discuss elsewhere \cite{examples}. The thermal circles for Rindler space and its T-dual partner are illustrated in figure \ref{thermalcircles}.
We conclude that in Euclidean Rindler space, there is no regime in which we can get control over the $\alpha'$ corrections to the thermal scalar field theory action if these terms are present in the first place, i.e. if their coefficients are non-zero. However, we can follow a different path and consider exact $SL(2,\mathbb{R})/U(1)$ WZW models that have a flat space limit corresponding to Euclidean Rindler space. This will be done in the next subsection.

\subsection{Exact WZW analysis}
In what follows we will rescale the time coordinate to $\tau = \frac{\sqrt{\alpha'}t}{4GM}$ to simplify several expressions and to make the link with $\mathbb{C}/\mathbb{Z}_N$ orbifolds more transparent. We have introduced the string length scale here as our reference length:
\begin{equation}
\label{snorm}
ds^2 = \left(\frac{\rho^2}{\alpha'}\right)d\tau^2 + d\rho^2 + d\mathbf{x}^2_{\perp}.
\end{equation}
To avoid a conical singularity, the Euclidean time coordinate needs to have the identification $\tau \sim \tau + \beta_R$ where $\beta_R = 2\pi \sqrt{\alpha'}$. \\
Let us now consider the $SL(2,\mathbb{R})/U(1)$ cigar CFT. This CFT was first discovered by Witten in \cite{Witten:1991yr} and has received a lot of attention since then (see e.g. \cite{Dijkgraaf:1991ba}\cite{Tseytlin:1991ht}\cite{Jack:1992mk}\cite{Giveon:1999px}\cite{Kutasov:2000jp}\cite{Kutasov:2005rr}\cite{Sugawara:2012ag}). To lowest order in $\alpha'$ the solution is given by
\begin{align}
\label{lowest}
ds^2 &= \frac{\alpha'k}{4}\left(dr^2 + 4\tanh^2\left(\frac{r}{2}\right)d\theta^2\right), \\
\label{lowest2}
\Phi &= - \ln\cosh\left(\frac{r}{2}\right),
\end{align}
where $\theta \sim \theta +2\pi$. Excitations in this background were studied extensively by \cite{Dijkgraaf:1991ba}.
The lowest lying state (tachyon) has the property
\begin{equation}
\label{ons}
(L_{0} + \bar{L}_0 - 2) \left|T\right\rangle = 0.
\end{equation}
For the cigar gauged WZW model, we can write this as a differential equation using the Laplacian on the group manifold (see \cite{Dijkgraaf:1991ba} for details):
\begin{align}
L_{0} &= -\frac{\Delta}{k-2} - \frac{1}{k}\partial_{\theta_L}^2, \\
\bar{L}_{0} &= -\frac{\Delta}{k-2} - \frac{1}{k}\partial_{\theta_R}^2,
\end{align}
where $\Delta$ denotes the Laplacian on the $SL(2,\mathbb{R})$ manifold. 
The physical state condition 
\begin{equation}
(L_0 - \bar{L}_0) \left|T\right\rangle = 0,
\end{equation} 
implies that tachyons divide in two categories for spaces where the other dimensions do not allow states with $L^{other}_{0} \neq \bar{L}^{other}_{0}$: momentum modes and winding modes. We are however interested in the one-loop free energy in which off-shell stringy states propagate in the loop. This implies we do not have to impose this physicality condition but instead we can relax it to $L_0 - \bar{L}_0 \in \mathbb{Z}$ which is required for modular invariance. For simplicity we will consider only pure winding and pure discrete momentum modes for now, though we will generalize this further on. 
The momentum modes have the following $L_0$:
\begin{equation}
L_{0} = - \frac{1}{k-2}\left[\partial_r^2 + \coth (r) \partial_r + \frac{1}{4}\left(\coth^2\left(\frac{r}{2}\right) - \frac{2}{k}\right)\partial_{\theta}^2\right].
\end{equation}
For winding modes on the other hand, the relevant operator is
\begin{equation}
L_{0} = - \frac{1}{k-2}\left[\partial_r^2 + \coth (r) \partial_r + \frac{1}{4}\left(\tanh^2\left(\frac{r}{2}\right) - \frac{2}{k}\right)\partial_{\tilde{\theta}}^2\right].
\end{equation}
For more details regarding these constructions, the reader is referred to \cite{Dijkgraaf:1991ba}.
The geometry can then be identified by writing: 
\begin{equation}
\label{metricL0}
L_0 = -\frac{\alpha'}{4e^{-2\Phi}\sqrt{G}}\partial_i e^{-2\Phi}\sqrt{G}G^{ij}\partial_j,
\end{equation}
since the on-shell relation (\ref{ons}) should contain the same content as the equations of motion for this field. This identifies immediately the effective metric and dilaton for the momentum modes as
\begin{align}
\label{momgeom}
ds^2 &= \frac{\alpha'}{4}(k-2)\left[dr^2 + \frac{4}{\coth^2\left(\frac{r}{2}\right) - \frac{2}{k}}d\theta^2\right],\\
\Phi &= -\frac{1}{2}\ln\left(\frac{\sinh(r)}{2}\sqrt{\coth^2\left(\frac{r}{2}\right) - \frac{2}{k}}\right),
\end{align}
and to lowest order in $1/k$ these agree with (\ref{lowest}) and (\ref{lowest2}). For the winding modes on the other hand, one finds
\begin{align}
\label{windgeom}
ds^2 &= \frac{\alpha'}{4}(k-2)\left[dr^2 + \frac{4}{\tanh^2\left(\frac{r}{2}\right) - \frac{2}{k}}d\tilde{\theta}^2\right],\\
\Phi &= -\frac{1}{2}\ln\left(\frac{\sinh(r)}{2}\sqrt{\tanh^2\left(\frac{r}{2}\right) - \frac{2}{k}}\right).
\end{align}
In this case the coordinate is identified as $\tilde{\theta} \sim \tilde{\theta} + \frac{2\pi}{k}$. Note that this identification is \emph{not} the same as the T-dual periodicity. The T-dual coordinate would need to have the periodicity $\theta^{Tdual} \sim \theta^{Tdual} + \frac{2\pi}{k-2}$.
This method of obtaining the background metric and dilaton proved to be very powerful for gauged WZW models \cite{Bars:1992sr}. In \cite{Dijkgraaf:1991ba} it was argued that the metric and dilaton obtained in this way are exact in $\alpha'$ and later on substantial evidence for this appeared \cite{Tseytlin:1991ht}\cite{Jack:1992mk}. This method also works for other coset models (see e.g. \cite{Bars:1992sr}).
These two backgrounds are \emph{not} related by the normal T-duality as discussed in \cite{Dijkgraaf:1991ba}. What does this mean? The momentum geometry (\ref{momgeom}) is what non-stringy excitations experience and this is the exact form: the Laplacian in this background determines the momentum modes. Performing a naive T-duality on the discrete momentum tachyon action, we obtain a winding tachyon action whose geometry does not correspond to the above exact dual geometry (\ref{windgeom}). This means that we can regard the mismatch as higher $\alpha'$ corrections to the T-dual tachyon action. When summing all of the $\alpha'$ corrections, we can rewrite the winding tachyon action as only a covariant kinetic term in the above exact dual background. This implies the winding tachyon action indeed gets $\alpha'$-corrected in this case.\\
For type II superstrings however, the story is different. The $L_0$ operator for momentum states is now
\begin{equation}
L_{0} = - \frac{1}{k}\left[\partial_r^2 + \coth (r) \partial_r + \frac{1}{4}\coth^2\left(\frac{r}{2}\right)\partial_{\theta}^2\right].
\end{equation}
For winding modes on the other hand, the relevant operator is
\begin{equation}
L_{0} = - \frac{1}{k}\left[\partial_r^2 + \coth (r) \partial_r + \frac{1}{4}\tanh^2\left(\frac{r}{2}\right)\partial_{\tilde{\theta}}^2\right].
\end{equation}
The overall normalization changed, but also the (innocent-looking) $\frac{2}{k}$ term in the $\theta$ derivatives disappeared. This is crucial since it is this last term that disrupted the interpretation of the dual background as just the T-dualized version. In this case the momentum modes perceive the metric and dilaton background as
\begin{align}
\label{metricn}
ds^2 &= \frac{\alpha' k}{4}\left[dr^2 + \frac{4}{\coth^2\left(\frac{r}{2}\right)}d\theta^2\right],\\
\label{diln}
\Phi &= -\ln\left(\cosh\left(\frac{r}{2}\right)\right),
\end{align}
and for the winding modes one finds
\begin{align}
\label{metric}
ds^2 &= \frac{\alpha' k}{4}\left[dr^2 + \frac{4}{\tanh^2\left(\frac{r}{2}\right)}d\tilde{\theta}^2\right],\\
\label{dil}
\Phi &= -\ln\left(\sinh\left(\frac{r}{2}\right)\right).
\end{align}
The coordinate periodicity is again $\theta \sim \theta +2\pi$ and $\tilde{\theta} \sim \tilde{\theta} + \frac{2\pi}{k}$ and now T-duality is manifest. This implies the winding tachyon action is obtained by simply T-dualizing the momentum tachyon action and this contains only the covariant kinetic term. The mass term finally is inserted because the on-shell conditions read\footnote{This only holds for $h = \bar{h}$ states.}
\begin{align}
(L_{0} - 1) \left|T\right\rangle = 0, \quad \text{bosonic},\\
(L_{0} - 1/2) \left|T\right\rangle = 0, \quad \text{type II}.
\end{align}
We conclude that for the type II string on the cigar, the naive lowest order $\alpha'$ effective tachyon action is exact. The dual metric (\ref{metric}) and dilaton (\ref{dil}) are in this case equal to the T-dual metric and dilaton of (\ref{metricn}) and (\ref{diln}) of the exact $\alpha'$ geometry. We interpret the dual metric and dilaton as a way to succinctly write down the winding tachyon action. Expanding this dual background around the T-dual background allows us to identify the $\alpha'$ corrections to the winding tachyon action. Since in this case, the dual and T-dual backgrounds are equal, no $\alpha'$ corrections are present.\\
So far, everything we discussed concerns the cigar CFT. Let us now consider the limit $k \to \infty$ to get to Euclidean Rindler space as discussed in \cite{Giveon:2013ica}. For type II superstrings, the authors of \cite{Giveon:2013ica} show that in this case the lowest order $\alpha'$ action is obtained by using this limit on the winding tachyon action in the cigar background. Since we saw above that this action should not get any corrections in $\alpha'$, the Rindler space winding tachyon action should also not get any corrections, at least if $k \to \infty$ is a good description of Rindler space. We conclude that the type II superstring lowest order effective winding tachyon action on Rindler space is exact.\\
Let us analyze bosonic strings more closely. As a first step, we rescale the dual angular coordinate $\tilde{\theta}$ as $\tilde{\theta}_{new} = \frac{k}{k-2}\tilde{\theta}_{old}$. The reason for doing this is that we now have a natural relation between $\theta$ and $\tilde{\theta}$ in terms of T-dual variables. Next we rescale the coordinates as $\rho = \frac{\sqrt{\alpha' (k-2)}}{2} r$, $\tau = \sqrt{\alpha'(k-2)}\theta$ and $\tilde{\tau} = \sqrt{\alpha'(k-2)}\tilde{\theta}$ in the exact bosonic cigar background and its dual. This gives
\begin{align}
ds^2 &= d\rho^2 + \frac{d\tau^2}{\coth^2\left(\frac{\rho}{\sqrt{\alpha'(k-2)}}\right) - \frac{2}{k}}, \quad \text{momentum metric},\\
\label{windmet}
ds^2 &= d\rho^2 + \left(\frac{k-2}{k}\right)^2\frac{d\tilde{\tau}^2}{\tanh^2\left(\frac{\rho}{\sqrt{\alpha'(k-2)}}\right) - \frac{2}{k}}, \quad \text{winding metric}.
\end{align}
where we now require $ \tau \sim \tau + 2\pi\sqrt{\alpha'(k-2)}$ and $\tilde{\tau} \sim \tilde{\tau} + 2\pi\sqrt{\frac{\alpha'}{k-2}}$. Note that indeed $\tau$ and $\tilde{\tau}$ have the correct T-dual periodicities. Let us call $\sqrt{\alpha'(k-2)} = L$. The winding tachyon action in the dual geometry is given by (corresponding to (\ref{ons}))
\begin{align}
&\int_{0}^{+\infty}d\rho\frac{L}{2}\sinh\left(\frac{2}{L}\rho\right)\left[\left|\partial_\rho T\right|^2 + w^2\tilde{G}^{00}\frac{\beta^2}{4\pi^2\alpha'^2}TT^{*} - \frac{4}{\alpha'}TT^{*}\right] \\
&= \int_{0}^{+\infty}d\rho\frac{L}{2}\sinh\left(\frac{2}{L}\rho\right)\left[\left|\partial_\rho T\right|^2 + w^2\frac{\beta^2}{4\pi^2\alpha'^2}\frac{k^2}{(k-2)^2}\left(\tanh^2\left(\frac{\rho}{L}\right) - \frac{2}{k}\right)TT^{*} - \frac{4}{\alpha'}TT^{*} \right]
\end{align}
and this action is exact in $\alpha'$ when considering $\beta$ equal to the inverse Hawking temperature on the cigar.\\
We follow the prescription of \cite{Giveon:2013ica} and let $k \to \infty$ in this action keeping $\rho$ fixed. We obtain
\begin{equation}
\int_{0}^{+\infty}d\rho\rho\left[\left|\partial_\rho T\right|^2 + w^2\frac{\beta^2\rho^2}{4\pi^2\alpha'^3(k-2)}TT^{*} - 2w^2\frac{\beta^2}{4\pi^2\alpha'^2k}TT^{*} - \frac{4}{\alpha'}TT^{*} \right].
\end{equation}
To further ease the identification with the string-normalized Rindler space discussed above in (\ref{snorm}), we extract a $\sqrt{k-2}$ factor from $\beta$ such that the canonical temperature becomes $\beta = 2\pi\sqrt{\alpha'}$ instead of $\beta = 2\pi\sqrt{\alpha'(k-2)}$. After doing this, we obtain for the $w=\pm1$ thermal scalar in the $k\to\infty$ limit:
\begin{equation}
\label{bosts}
\int_{0}^{+\infty}d\rho\rho\left[\left|\partial_\rho T\right|^2 + \frac{\beta^2\rho^2}{4\pi^2\alpha'^3}TT^{*} - 2\frac{\beta^2}{4\pi^2\alpha'^2}TT^{*} - \frac{4}{\alpha'}TT^{*} \right].
\end{equation}
Note that the metric component $\tilde{G}_{00}$ in equation (\ref{windmet}) has a component $\propto 1/k$ which would at first sight vanish when taking the $k\to\infty$ limit. This would leave the large $k$ action unaltered w.r.t. the lowest order (in $\alpha'$) thermal scalar action. However, $\beta^2$ is also proportional to $k$ for large $k$. Hence these factors cancel and leave a finite contribution in the limit. 
Plugging in the canonical Rindler temperature finally yields:
\begin{equation}
\int_{0}^{+\infty}d\rho\rho\left[\left|\partial_\rho T\right|^2 + \frac{\rho^2}{\alpha'^2}TT^{*} - \frac{2}{\alpha'}TT^{*} - \frac{4}{\alpha'}TT^{*} \right].
\end{equation}
We conclude that the only effect of all other higher order $\alpha'$ corrections is a mass shift. Such terms were indeed expected to correct the field theory action as we discussed in the previous subsection.\\
To summarize, for Euclidean Rindler space the bosonic string thermal scalar action receives $\alpha'$ corrections as determined above, whereas the type II superstring thermal scalar action does not get $\alpha'$-corrected.

\section{Critical behavior in Rindler space}
\label{critical}
Using the above field theory actions for the thermal scalar, we can analyze the critical behavior of the one-loop free energy using (\ref{FT}). We can then use (\ref{randwalk}) to rewrite this expression as a random walk in the purely spatial submanifold. Note that we do not yet know what `critical' means for Rindler space, since we have not yet determined the Hagedorn temperature. This will also be done using the thermal scalar action.
In this section we discuss the thermal scalar approximation to the free energy for type II, heterotic and bosonic strings, where we keep the bosonic strings for last since these present (surprisingly!) the most subtleties.

\subsection{Type II Superstrings in Rindler space}
\label{typeII}
We first consider type II superstrings since these do not present complications in the spectrum and these are also more realistic than bosonic strings. The one-loop free energy of the thermal scalar field is given by
\begin{equation}
\label{freebosonic}
\beta F = - \int_{0}^{+\infty}\frac{dT}{T} \text{Tr} e^{-T\left(-\nabla^{2} + m_{local}^2 - G^{ij}\frac{\partial_{j}\sqrt{G_{00}}}{\sqrt{G_{00}}}\partial_{i}\right)}.
\end{equation}
For Rindler spacetime with a general $\beta$ (flat conical spaces), the operator in the exponential is given by
\begin{equation}
\label{opera}
-\partial^{2}_{\rho} - \frac{1}{\rho}\partial_{\rho} - \frac{2}{\alpha'} + \frac{\beta^2\rho^2}{4\pi^2\alpha'^3}.
\end{equation}
We now search for the eigenfunctions and eigenvalues of this operator. Enforcing regularity at the origin and at infinity gives a discrete set of eigenfunctions and eigenvalues given by
\begin{align}
\label{eigenf}
\psi_{n}(\rho) &\propto \exp\left(-\frac{\beta\rho^2}{4\pi\alpha'^{3/2}}\right)L_{n}\left(\frac{\beta\rho^2}{2\pi\alpha'^{3/2}}\right), \\
\label{superspectrum}
\lambda_{n} &= \frac{\beta - 2\pi\sqrt{\alpha'}+2\beta n}{\pi \alpha'^{3/2}},
\end{align}
where $L_{n}$ is the Laguerre polynomial and $n$ is a positive (or zero) integer. The $n=0$ term has the lowest eigenvalue. Setting $\beta = 2\pi\sqrt{\alpha'}$ and $n=0$ gives
\begin{equation}
\label{gs}
\psi_0 \propto \exp\left(-\frac{\rho^2}{2\alpha'}\right), \quad \lambda_0=0.
\end{equation}
The large $T$ contribution is then given by
\begin{equation}
\label{meth22}
\text{Tr} e^{-T\left(-\nabla^{2} + m_{local}^2 - G^{ij}\frac{\partial_{j}\sqrt{G_{00}}}{\sqrt{G_{00}}}\partial_{i}\right)} \to e^{-\lambda_{0}T} = \exp\left(-\frac{\beta - 2\pi\sqrt{\alpha'}}{\pi \alpha'^{3/2}}T\right).
\end{equation}
Let us reinterpret this one-loop free energy as a random walk in the spatial submanifold as we did in going from equation (\ref{FT}) to (\ref{randwalk}).
Applying formula (\ref{randwalk}) gives
\begin{equation}
\beta F = - \int_{0}^{+\infty}\frac{dT}{T}\int_{S^{1}} \left[\mathcal{D}x\right]\sqrt{G_{sp}}e^{-S - S_{sp}}
\end{equation}
where
\begin{equation}
\label{ppp2}
S = \frac{1}{4\pi\alpha'}\int_{0}^{T}dt\left[\dot{\rho}^2+\frac{\beta^2\rho^2}{\alpha'}-\frac{\pi^2\alpha'^2}{\rho^2}-8\pi^2\alpha' \right]
\end{equation}
and $G_{sp}$ and $S_{sp}$ denote the metric and action of the other spectator dimensions needed to get a valid string background. We are not interested in these at the moment.\\
Our original goal was to get information on the corrections to the worldsheet dimensional reduction approach as discussed in \cite{theory}. Let us look at the particle action (\ref{ppp2}) more closely. The first two terms are the only ones that are found in the Fourier expansion in the string path integral as discussed in \cite{theory}. The third term was denoted $K(x)$ in section \ref{pathderiv} and it is caused by removing the $\sqrt{G_{00}}$ from the measure of the field theory action as we discussed above. The fourth term is the Hagedorn correction that can be found by looking at flat space models. The free energy corresponding to the above action (\ref{ppp2}) has a random walk interpretation with a modified potential. \\
The result is a particle path integral on a half-line in a harmonic oscillator and in a $1/\rho^2$ potential. From equation (\ref{meth22}) we can distill the Hagedorn temperature. To do this, of course, we need to know what the other spectator dimensions are. These can obviously influence the Hagedorn temperature. Let us choose 24 flat dimensions such that the total one-loop free energy of the thermal scalar becomes
\begin{equation}
\label{type2free}
\beta F = -V_{T}\int_{0}^{+\infty}\frac{dT}{T}\left(\frac{1}{4\pi T}\right)^{12}\exp\left(-\frac{\beta}{\pi\alpha'^{3/2}}T + \frac{2}{\alpha'}T \right).
\end{equation}
Crucially, the traced heat kernel of flat dimensions do not yield corrections to the exponential. Convergence in the large $T$ limit then determines the critical temperature as:
\begin{equation}
\label{convsuper}
\beta \geq 2\pi\sqrt{\alpha'}.
\end{equation}
We clearly find $\beta_{critical} = \beta_{R}$ so the canonical Rindler temperature (needed to avoid the conical singularity and defined in equation (\ref{snorm})) is precisely equal to the critical temperature above which the free energy would diverge in the IR. We will comment on the link to the physical black hole normalized coordinates (\ref{Rindmetric}) further on. This marginal convergence is associated with a state in the string spectrum that becomes massless precisely when $\beta = 2\pi\sqrt{\alpha'}$ and this stringy state was found by the authors of \cite{Kutasov:2000jp}\cite{Giveon:2012kp}\cite{Giveon:2013ica}. \\
For any $\beta$, the wavefunction of the lowest state ($n=0$) is given by
\begin{equation}
\label{groundstate}
\psi_0 \propto \exp\left(-\frac{\beta\rho^2}{4\pi\alpha'^{3/2}}\right).
\end{equation}
We identify the lowest eigenfunction of the associated particle heat kernel with the wavefunction of the thermal scalar stringy state. So qualitatively the type II thermal scalar in Rindler space is a massless state located close to the origin in Euclidean Rindler space. As a sidenote, we remark that expression (\ref{type2free}) is also UV divergent, although this is a remnant of considering solely the thermal scalar field theory and not the full string theory. Ultimately we should not take $0$ as the lower boundary of the $T$-integral since we assumed $T$ large. This is actually the same story as in flat space: the difference between the fundamental modular domain and the strip is the reason for the divergence.\\

Note also that even though the original string path integral was only well-defined for $\beta = \frac{2\pi\sqrt{\alpha'}}{N}$ (for $N \neq 1$ these models are flat space $\mathbb{C}/\mathbb{Z}_N$ orbifolds that have been used in the past to predict string thermodynamics \cite{Dabholkar:1994ai}\cite{Lowe:1994ah} and we will discuss these further on), the resulting particle path integral and the final result have been calculated for a general $\beta$. Thus this provides a \emph{natural} continuation of the string results to a general $\beta$. The continuation is in terms of the field theory of the states constituting the string theory, and this road to off-shell descriptions of string theory has indeed been the most fruitful one (see e.g. \cite{Taylor:2003gn} and references therein). In particular, this allows us to differentiate with respect to $\beta$ to obtain the thermodynamic entropy.\\

The authors of \cite{Kruczenski:2005pj} give a formula for the average size of the long string. They showed that it is given by the width of the ground state wavefunction of the associated particle heat kernel. Applying such a reasoning to the ground state wavefunction given in (\ref{groundstate}), we find as a measure for the width of the $n=0$ mode that
\begin{equation}
\left\langle \rho^2\right\rangle = \frac{2\pi\alpha'^{3/2}}{\beta}.
\end{equation}
Note that this assumes that values of $\beta \neq 2\pi \sqrt{\alpha'}$ are meaningful as we have discussed above.\\

To sum up, the wavefunction of the thermal scalar is identified with the lowest eigenfunction of the particle heat kernel and this determines the region where the random walk is situated. Thus the picture we arrive at is that the thermal scalar is represented by a random walk close to the Rindler origin. This interpretation is the same as in flat spacetime, the difference is that in Rindler space the walk is localized close to the origin. Since the Rindler origin is actually the black hole horizon, we conclude that the thermal scalar is localized to a string length surrounding a black hole. For clarity about the transition from this string-normalized Rindler spacetime (\ref{snorm}) to the black hole-normalized Rindler spacetime (\ref{Rindmetric}), we refer to appendix \ref{GM} where we translate the results from this section to the black hole case.

\subsection{Heterotic strings in Rindler space}
\label{heterotic}
We can also solve the previous model for the heterotic string in Rindler spacetime. In this case, we do not at first sight have a WZW cigar model to guide us, so we will compute the critical behavior \emph{assuming} no $\alpha'$ corrections to the thermal scalar action, just like for the type II case. We will provide arguments in favor of this further on. The operator $\hat{\mathcal{O}}$ to be considered for the heterotic string is given by
\begin{equation}
\frac{1}{2}\left(-\partial^{2}_{\rho} - \frac{1}{\rho}\partial_{\rho} - \frac{3}{\alpha'} + \frac{\beta^2\rho^2}{4\pi^2\alpha'^3} + \frac{\alpha' \pi^2}{\beta^2 \rho^2}\right),  
\end{equation}
with eigenfunctions and eigenvalues given by
\begin{align}
\psi_{n}(\rho) \propto \rho^{\frac{\pi\sqrt{\alpha'}}{\beta}}\exp\left(-\frac{\beta\rho^2}{4\pi\alpha'^{3/2}}\right)L_{n}^{\left(a\right)}\left(\frac{\beta\rho^2}{2\pi\alpha'^{3/2}}\right), \quad \lambda_{n} = \frac{\beta - 2\pi\sqrt{\alpha'}+2\beta n}{\pi \alpha'^{3/2}},
\end{align}
where in this case the generalized Laguerre polynomial is used with order $a=\frac{\pi\sqrt{\alpha'}}{\beta}$. The lowest eigenfunction is given by
\begin{equation}
\psi_0 \propto \rho^{\frac{\pi\sqrt{\alpha'}}{\beta}}\exp\left(-\frac{\beta\rho^2}{4\pi\alpha'^{3/2}}\right).
\end{equation}
The ground state has again zero eigenvalue for the canonical Rindler temperature so, like in the type II case, the convergence criterion is given by:\footnote{When we include some extra flat dimensions.}
\begin{equation}
\label{convhet}
\beta \geq 2\pi \sqrt{\alpha'}.
\end{equation}
In particular we have again that the canonical Rindler temperature equals the Rindler Hagedorn temperature. At this temperature, the zero-mode has the wavefunction
\begin{equation}
\psi_0 \propto \rho^{\frac{1}{2}}\exp\left(-\frac{\rho^2}{2\alpha'}\right), \quad \lambda_0=0.
\end{equation}
For the heterotic string, the $n=0$ width formula changes to 
\begin{equation}
\left\langle \rho^2\right\rangle = \frac{2\pi\alpha'^{3/2}}{\beta}\left(1+\frac{\pi\sqrt{\alpha'}}{\beta}\right).
\end{equation} 
In particular, for the canonical Rindler temperature, we get a size equal to $\sqrt{\frac{3}{2}}\sqrt{\alpha'}$ which is a factor of $\sqrt{\frac{3}{2}}$ \emph{larger} than the case considered above.\\
The random walk behavior has the form:
\begin{equation}
\beta F = - \int_{0}^{+\infty}\frac{dT}{T}\int_{S^{1}} \left[\mathcal{D}x\right]\sqrt{G_{sp}}e^{-S - S_{sp}}
\end{equation}
where
\begin{equation}
S = \frac{1}{4\pi\alpha'}\int_{0}^{T}dt\left[\dot{\rho}^2+\frac{\beta^2\rho^2}{\alpha'} + \frac{4\pi^4\alpha'^3}{\beta^2\rho^2}-\frac{\pi^2\alpha'^2}{\rho^2}-12\pi^2\alpha' \right].
\end{equation}
Finally note that we have lost the heterotic thermal duality \cite{O'Brien:1987pn} in this case. The duality symmetry is compromised as soon as one considers a non-trivial background.

\subsection{Bosonic strings in Rindler space}
\label{bosonic}
In this section we discuss the same story for bosonic strings. This case is more complex due to the $\alpha'$ corrections and also due to unitarity constraints that we will discuss further on in section \ref{unitarity}. \\
If we include the $\alpha'$ corrections, we need to consider the operator
\begin{equation}
-\partial^{2}_{\rho} - \frac{1}{\rho}\partial_{\rho} - 2\frac{\beta^2}{4\pi^2\alpha'^2} - \frac{4}{\alpha'} + \frac{\beta^2\rho^2}{4\pi^2\alpha'^3},
\end{equation}
where also the substitution $\frac{2}{\alpha'} \to \frac{4}{\alpha'}$ was made in comparison to the type II superstring. The eigenfunctions remain the same as in the type II case but the eigenvalues shift to
\begin{equation}
\label{bosonicspectrum}
\lambda_{n} = \frac{\beta - 4\pi\sqrt{\alpha'} - \frac{\beta^2}{2\pi\sqrt{\alpha'}}+2\beta n}{\pi \alpha'^{3/2}}.
\end{equation}
The $n=0$ term has the lowest eigenvalue. This state has the same wavefunction and hence also the same width as the type II case discussed above. A further subtlety is whether these quantum numbers are really in the string spectrum. We will discuss this further in section \ref{unitarity}, where we will conclude that actually the bosonic spectrum only starts at $n=1$. In this section we will ignore this complication because other thermodynamic quantities (like the entropy) do appear to rely on the $n=0$ mode. Setting $\beta = 2\pi\sqrt{\alpha'}$ gives a negative $n=0$ eigenvalue. 
The (naive) critical behavior is given by
\begin{equation}
\label{meth1alpha}
\text{Tr} e^{-T\left(-\nabla^{2} + m_{local}^2 - G^{ij}\frac{\partial_{j}\sqrt{G_{00}}}{\sqrt{G_{00}}}\partial_{i}\right)} \to e^{-\lambda_{0}T} = \exp\left(-\frac{\beta - 4\pi\sqrt{\alpha'}-\frac{\beta^2}{2\pi\sqrt{\alpha'}}}{\pi \alpha'^{3/2}}T\right).
\end{equation}
The random walk interpretation can again be found by applying formula (\ref{randwalk}) and including the $\alpha'$ correction term in this case gives
\begin{equation}
\beta F = - \int_{0}^{+\infty}\frac{dT}{T}\int_{S^{1}} \left[\mathcal{D}x\right]\sqrt{G_{sp}}e^{-S - S_{sp}}
\end{equation}
where
\begin{equation}
\label{pppp}
S = \frac{1}{4\pi\alpha'}\int_{0}^{T}dt\left[\dot{\rho}^2+\frac{\beta^2\rho^2}{\alpha'}-\frac{\pi^2\alpha'^2}{\rho^2}-16\pi^2\alpha'- 2\beta^2\right]
\end{equation}
with $G_{sp}$ and $S_{sp}$ the metric and action of the spectator dimensions. \\
For bosonic strings in Rindler space, we have a nice demonstration of all different types of corrections to the naive particle action (\ref{act}). 
All terms have the same origin as in the type II case, except the final term. This term combines all $\alpha'$ corrections to the field theory action for this particular background. We saw previously that in general such terms were to be expected in Rindler space and the cigar CFT approach indeed produces such a correction term. We believe that now all corrections are determined. The free energy corresponding to the above action (\ref{pppp}) has a random walk interpretation with a modified potential and with a temperature-dependent mass. \\
As a further check on these results and in particular the manipulations done to get from (\ref{FT}) to (\ref{randwalk}), we explicitly solve the particle path integral directly and get the same results as above. We present these results in appendix \ref{appB}, where we also discuss the modifications needed to treat superstrings and heterotic strings. \\
Again choosing 24 flat dimensions, the total one-loop free energy of the thermal scalar becomes
\begin{equation}
\label{bosfree}
\beta F = -V_{T}\int_{0}^{+\infty}\frac{dT}{T}\left(\frac{1}{4\pi T}\right)^{12}\exp\left(-\frac{\beta}{\pi\alpha'^{3/2}}T + \frac{4}{\alpha'}T + \frac{\beta^2}{2\pi^2\alpha'^2}T\right).
\end{equation}
The last term in the exponential factor is the result of the $\alpha'$ corrections and its influence is substantial. If we ignored the $\alpha'$ corrections, we would determine the convergence criterion to be
\begin{equation}
\beta \geq 4\pi\sqrt{\alpha'}.
\end{equation}
If we include this final term however, we would find a divergence in thermodynamical quantities for any value of $\beta$:\footnote{Note that, as we previously discussed, we should think of this divergence as occuring in the entropy and not in the free energy itself since this quantity only starts with the $n=1$ mode as we will discuss further on.} the Hagedorn temperature is effectively zero (there is always a negative eigenvalue). \\
If we would (naively) consider arbitrary winding modes, it is readily checked that not only the $w=\pm1$ mode, but also all higher winding modes are tachyonic at the canonical Rindler temperature. This behavior is a drastic departure from the lowest $\alpha'$ action discussed above (i.e. ignoring the last term in the exponential in (\ref{bosfree})) where $w=\pm2$ is massless and all higher winding modes would be massive. This discussion is a bit mute since we will see that all higher winding modes are actually not in the Euclidean Rindler spectrum in the first place, but it is an illustration of the importance of the $\alpha'$ corrections in a cigar-shaped background (or its flat Euclidean Rindler limit considered here). \\

To conclude the bosonic thermal scalar in Rindler space, the lowest $n=0$ state is tachyonic and is situated at a string length from the origin of Rindler space. However, we will see in section \ref{unitarity} that the string spectrum only starts at $n=1$ in this case. We will postpone any further discussion on this until then. 

\subsection{Hagedorn behavior of the Rindler string}
Above we established the critical temperatures for the string gas in Rindler space. We found that for type II and heterotic strings, the free energy is `marginally' convergent. In this section we want to elaborate on whether the free energy converges or diverges at this critical temperature. So we zoom in on the behavior of the free energy at the Hagedorn temperature. The crucial aspect is whether the spectator dimensions are compact or non-compact. Let us first take the other dimensions (tangential to the horizon) as compact dimensions. We have to do this to circumvent the Jeans instability. The effect of compact dimensions on the Hagedorn behavior is illustrated in appendix \ref{dom}. For type II superstrings and for heterotic strings we obtain the following free energy expression
\begin{equation}
F = -\frac{V_{T}}{\beta}\int_{0}^{+\infty}\frac{dT}{T},
\end{equation}
which is analogous to (\ref{bosfree}) by replacing the non-compact dimensions by compact ones and adjusting for the superstring or heterotic string. We clearly see a logarithmic divergence for $T \to \infty$ in this case. The free energy at the Hagedorn temperature (which is the same as the canonical Rindler temperature) diverges in the IR and this divergence is caused by the massless states. Other non-winding massless states also cause this same kind of divergence, but they are independent of the temperature and their contribution is dropped (and these do not influence the entropy). We find that the thermal scalar yields the divergent temperature-dependent contribution to thermodynamic quantities in Rindler space and it should take over the entire thermodynamics. One can compare this type of divergence with the microcanonical picture in flat space where one can never reach the Hagedorn temperature and the free energy diverges at this temperature. This does not cause a condensation process, but represents here our inability to reach this temperature. The difference with this case and the Rindler case, is that here the temperature should remain fixed at the canonical Rindler temperature and this equals the Hagedorn temperature. So in some sence, the Rindler string is held fixed at this unobtainable temperature of strings-in-a-box.
If we only take the contribution from the thermal scalar, we find a random walk behavior as was predicted by Susskind.\\

Since $F/V_T$ becomes infinite, the string density becomes arbitrarily high. It is in this way that the higher order string interactions should come in and cure this behavior. In \cite{Susskind:2005js} it was argued that interactions would cause a repulsion in such a way that the density becomes constant when nearing the horizon. Higher genus near-Hagedorn thermodynamics has not been studied much in the past.\footnote{However see \cite{Brigante:2007jv} where the Hagedorn behavior is studied using factorizations of higher genus Riemann surfaces and dual holographic matrix models.} \\

If we would take at least one non-compact dimension, the free energy density becomes finite at the Hagedorn temperature. The thermal scalar no longer dominates the free energy. However, if we differentiate with respect to $\beta$, we bring powers of the Schwinger parameter $T$ down that deteriorate the convergence properties in the IR limit $T \to \infty$. In other words, the non-analyticity of the free energy gets its dominant contribution from the thermal scalar and the number of compact dimensions determines how many times one has to differentiate to obtain a divergence. This situation is exactly as in flat space. The behavior of the free energy near the Hagedorn temperature when including $D$ non-compact dimensions is given by
\begin{equation}
F \propto 
\left\{
    \begin{array}{ll}
        \left(\beta - \beta_{critical}\right)^{D/2}\ln\left(\beta - \beta_{critical}\right), \quad D\text{ even}, \\
        \left(\beta - \beta_{critical}\right)^{D/2}, \quad D\text{ odd}.
    \end{array}\right.
\end{equation}

We will now compute the asymptotic density of single string states and provide an interpretation of the long string and the stretched horizon. For type II superstrings and heterotic strings, the single string partition function is dominated by the zero-mode as given by (\ref{type2free}): 
\begin{equation}
Z = V_{T}\int_{0}^{+\infty}\frac{dT}{T}\left(\frac{1}{4\pi T}\right)^{D/2}\exp\left(-\frac{\beta}{\pi\alpha'^{3/2}}T + \frac{2}{\alpha'}T \right),
\end{equation}
where $V_T$ denotes the volume of the transverse dimensions, i.e. the area of the black hole horizon. We have chosen $D$ non-compact spectator dimensions in this discussion. The temperature $\beta$ in Rindler space is defined as the local temperature at $\rho = 4GM$ and given in string units corresponding with the definition of the Rindler metric as in (\ref{snorm}). In physical units where the metric is given by (\ref{Rindmetric}), this becomes
\begin{equation}
Z = V_{T}\int_{0}^{+\infty}\frac{dT}{T}\left(\frac{1}{4\pi T}\right)^{D/2}\exp\left(-\frac{\beta}{\pi\alpha'(4GM)}T + \frac{2}{\alpha'}T \right).
\end{equation}
Defining the energy at $\rho=4GM$ by 
\begin{equation}
E = \frac{T}{4\pi\alpha'GM},
\end{equation}
we find:
\begin{equation}
\label{star}
Z = V_{T}\int_{0}^{+\infty}\frac{dE}{E^{1+D/2}}\left(\frac{1}{16\pi^2\alpha'GM}\right)^{D/2}\exp\left(-\beta E + 8\pi GM E \right).
\end{equation}
The single string density of states is given by an inverse Laplace transform of this (single string) partition function  as
\begin{equation}
Z = \int_{0}^{+\infty}dE \rho(E) e^{-\beta E},
\end{equation}
which at high energies using (\ref{star}) yields
\begin{equation}
\rho(E) \approx V_T \left(\frac{1}{2\pi\beta_R \alpha'}\right)^{D/2}\frac{e^{\beta_R E}}{E^{1+D/2}}
\end{equation}
with $\beta_R = 8\pi GM$ the inverse Rindler temperature. The asymptotic density of states clearly displays Hagedorn behavior with critical temperature exactly equal to the Rindler temperature. Note also that the density of states is proportional to the transverse area (the horizon area). Let us now give an interpretation of the stretched horizon using these results. We can define the location of the long string (or the stretched horizon) as the distance from the horizon where the blueshifted canonical Rindler temperature becomes equal to the flat space Hagedorn temperature. Using the flat space Hagedorn expressions
\begin{align}
\beta_{H,flat} &= 2\sqrt{2}\pi\sqrt{\alpha'}, \quad \text{Type II}, \\
\beta_{H,flat} &= (2+\sqrt{2})\pi\sqrt{\alpha'}, \quad \text{Heterotic},
\end{align}
and
\begin{equation}
\beta_R \sqrt{G_{00}(\rho_{sh})} = \beta_R \frac{\rho_{sh}}{4GM} = \beta_{H,flat},
\end{equation}
where the subscript $sh$ denotes stretched horizon quantities, we can localize the stretched horizon at 
\begin{align}
\rho_{sh} &= \sqrt{2\alpha'}, \quad \text{Type II}, \\
\rho_{sh} &= \left(1 + \frac{1}{\sqrt{2}}\right) \sqrt{\alpha'}, \quad \text{Heterotic}.
\end{align}
The physical picture which arises from this, is that the thermodynamics of a gas of strings in a Rindler background is dominated by a single long string living at $\rho_{sh} \propto \sqrt{\alpha'}$ at the flat space Hagedorn temperature.\footnote{Again we should take into account the compactness of the other dimensions to determine whether one string dominates or multiple-string configurations dominate \cite{Deo:1989bv}.} The density of single string states as measured by a Rindler observer at the stretched horizon is then given by
\begin{equation}
\rho(E_{sh}) \approx V_T \left(\frac{1}{2\pi\beta_R \alpha'}\right)^{D/2}\left(\frac{\beta_R}{\beta_{H,flat}}\right)^{D/2}\frac{e^{\beta_{H,flat} E_{sh}}}{E_{sh}^{1+D/2}},
\end{equation}
where we used the redshift $\beta_R E = \beta_{H,flat} E_{sh}$ and $\rho(E) dE = \rho(E_{sh}) dE_{sh}$. For a black hole, the asymptotic (single string) density of states is the same and the Rindler temperature equals the Hawking temperature in this case, as we have discussed already before. This behavior can be interpreted as Hagedorn critical behavior which is redshifted to the Hawking temperature for the observer at infinity.

\section{Comparison to flat $\mathbb{C}/\mathbb{Z}_N$ orbifold CFTs}
\label{orbifold}
\begin{figure}[h]
\centering
\includegraphics[width=7cm]{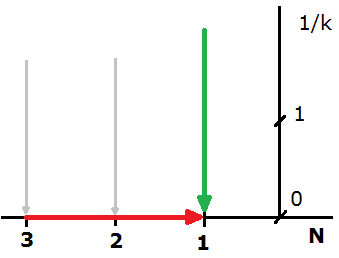}
\caption{$1/k$ versus orbifold number $N$. The horizonal red arrow represents the $\mathbb{C}/\mathbb{Z}_N$ continuation to $N=1$. The vertical green line represents the approach taken by the authors of \cite{Giveon:2013ica}, where one takes $k\to\infty$ in the cigar CFT. The vertical gray lines represent the strategy we are following in this section.}
\label{logic}
\end{figure}

In this section we will provide additional evidence for the recent claim by \cite{Giveon:2013ica} we discussed above that the $\alpha'$ corrections to the thermal scalar action can be deduced from the action on the $SL(2,\mathbb{R})/U(1)$ cigar. We will do this by comparing the resulting critical one-loop free energy with the one coming from flat space orbifolds \cite{Dabholkar:1994ai}\cite{Lowe:1994ah}. The orbifolds we have in mind are obtained by creating a conical space from a two-dimensional plane: the $\mathbb{C}/\mathbb{Z}_N$ orbifolds. We refer the reader to \cite{Dabholkar:1994ai}\cite{Lowe:1994ah} for a treatment of the spectrum and the partition function. These have been used as an approach to thermodynamics of the Rindler observer. Above we argued that we should be able to vary $\beta$ continuously to study thermodynamics. For string theory, only $\beta = \frac{2\pi\sqrt{\alpha'}}{N}$ correspond to CFTs, namely those obtained by orbifolding flat space. The spirit of \cite{Dabholkar:1994ai}\cite{Lowe:1994ah} is to take these discrete values of $\beta$ and afterwards analytically continue the resulting expression to a general $\beta$. Although the continuation and interpretation of these results to Rindler space ($N \to 1$) is not entirely airtight, for $\beta = \frac{2\pi\sqrt{\alpha'}}{N}$ the orbifold construction is well-founded in string theory and so it is an important check if we reproduce this with the cigar $k\to \infty$ approach. We will find perfect agreement. To clarify our intents, we sketch the strategy in figure \ref{logic}.
Taking $k\to\infty$ in the $SL(2,\mathbb{R})/U(1)$ model as in \cite{Giveon:2013ica}, we obtain a field theory action that \emph{supposedly} is the Rindler thermal scalar action. We do not have a verification for this result since the $\mathbb{C}/\mathbb{Z}_N$ continuation does not give a prediction for the thermal scalar action for $N=1$.\footnote{Simply setting $N=1$ in the $\mathbb{C}/\mathbb{Z}_N$ partition function of \cite{Dabholkar:1994ai}\cite{Lowe:1994ah} yields the partition function of an infinite 2d plane. This does not reflect the thermal scalar winding the Euclidean origin and is useless when we are interested in for instance the thermal entropy. This is also the reason Rindler thermodynamics is more complicated than simply stating that Rindler space is flat space.} For $N>1$, the $\mathbb{C}/\mathbb{Z}_N$ construction \emph{does} give a prediction for the thermal scalar action and we will find a precise match with this result. \\
Consider the bosonic string partition function on the $\mathbb{C}/\mathbb{Z}_N$ orbifold \cite{Dabholkar:1994ai}\cite{Lowe:1994ah}:
\begin{equation}
Z = \frac{V_{T}}{N}\int_{F}\frac{d\tau_1d\tau_2}{2\tau_2}\frac{1}{\left|\eta(\tau)\right|^{44}(4\pi^2\alpha'\tau_2)^{12}}\sum_{m,n=0}^{N-1}Z_{m,n}
\end{equation}
where 
\begin{equation}
Z_{m,n} = \left| \frac{\eta(\tau)}{\theta\left[
\begin{array}{c}
\frac{1}{2}+\frac{m}{N} \\
\frac{1}{2}+\frac{n}{N}  \end{array} 
\right](\tau)}\right|^2
\end{equation}
for $(m,n) \neq (0,0)$ and $Z_{0,0} = \frac{A}{\tau_2\left|\eta(\tau)\right|^4}$. The quantity $V_T$ denotes the volume of the other dimensions which have been chosen flat. The $m=n=0$ sector has $Z \propto \beta$ and so gives a temperature-independent contribution to the free energy and this is dropped from now on.
We study the $\tau_2 \to \infty$ limit in this case.\footnote{It is easier to take this limit in the fundamental modular domain.} 
The theta function with characteristics has the asymptotic behavior
\begin{equation}
\left|\theta\left[
\begin{array}{c}
\frac{1}{2} + \frac{m}{N} \\
b  \end{array} 
\right](\tau)\right| \to e^{-\pi \left(\frac{1}{2} + \frac{m}{N}\right)^2 \tau_2}e^{2\pi\tau_2\frac{m}{N}}.
\end{equation}
So we get\footnote{To be precise, the $m=0$ sector has a different prefactor since we dropped the $m=n=0$ sector. We do not care about this, since we will drop the $m=0$ sector anyway in what follows.}
\begin{equation}
Z = V_{T}\int^{+\infty}\frac{d \tau_2}{2\tau_2}\frac{1}{(4\pi^2\alpha'\tau_2)^{12}}\sum_{m=0}^{N-1}e^{2\pi\tau_2\left(2-\frac{m}{N}+\frac{m^2}{N^2}\right)}.
\end{equation}
The $\tau_1$ integral gives a multiplicative factor of 1. The $m=1$ and $m=N-1$ sectors dominate so the critical behavior is given by
\begin{equation}
Z = V_{T}\int^{+\infty}\frac{d \tau_2}{\tau_2}\frac{1}{(4\pi^2\alpha'\tau_2)^{12}}e^{2\pi\tau_2\left(2-\frac{1}{N}+\frac{1}{N^2}\right)}.
\end{equation}
The critical behavior of the free energy becomes
\begin{equation}
\label{free}
F = -\frac{V_{T}N}{2\pi}\int^{+\infty}\frac{d \tau_2}{\tau_2}\frac{1}{(4\pi^2\alpha'\tau_2)^{12}}e^{2\pi\tau_2\left(2-\frac{1}{N}+\frac{1}{N^2}\right)}.
\end{equation}
Looking back at the bosonic thermal scalar action (\ref{bosts}), we fill in the orbifold temperatures ($\beta = \frac{2\pi\sqrt{\alpha'}}{N}$) and obtain\footnote{This state actually coincides with the most divergent state in the $SL(2,\mathbb{R})/U(1)$ conical orbifolds as it should.}
\begin{equation}
\int_{0}^{+\infty}d\rho\rho\left[\left|\partial_\rho T\right|^2 + w^2\frac{\rho^2}{N^2\alpha'^2}TT^{*} - \frac{2}{\alpha'}\frac{w^2}{N^2}TT^{*} - \frac{4}{\alpha'}TT^{*} \right].
\end{equation}
The partition function becomes
\begin{equation}
Z = V_{T}\int_{0}^{+\infty}\frac{dT}{T}\left(\frac{1}{4\pi T}\right)^{12}e^{-\frac{2}{N\alpha'} T}e^{\frac{4}{\alpha'}T}e^{\frac{2}{N^2\alpha'}T}.
\end{equation}
Performing the substitution $T = \pi\alpha'\tau_2$, we obtain
\begin{equation}
\label{orbi}
F = -\frac{V_{T}N}{2\pi}\int_{0}^{+\infty}\frac{d\tau_2}{\tau_2}\left(\frac{1}{4\pi^2\alpha' \tau_2}\right)^{12}e^{2\pi\tau_2\left(2 - \frac{1}{N} + \frac{1}{N^2}\right)},
\end{equation}
which precisely coincides with (\ref{free}).\\
Next we discuss type II and heterotic superstrings. We will suffice by comparing the criterion for divergence of the partition function, although an elaborate treatment like for the bosonic string discussed above is possible. It is shown in \cite{Lowe:1994ah} that oddly twisted sectors (with twist $w$) have tachyons with masses $M^2 = \frac{2}{\alpha'}\left(\frac{w}{N}-1\right)$. Since with our notation $w\beta = \frac{w}{N}2\pi\sqrt{\alpha'}$, we find that the convergence criterion for the free energy that we found earlier in (\ref{convsuper}) and (\ref{convhet}) can be rewritten as
\begin{equation}
\label{convcrit}
0 \leq \beta - 2\pi \sqrt{\alpha'} = 2\pi\sqrt{\alpha'}\left(\frac{1}{N} - 1\right) = \pi \alpha'^{3/2}M^2.
\end{equation}
So we find that for $N \in \mathbb{N}$, the convergence criterion is equivalent to whether the most tachyonic twisted state has $M^2 > 0$ or $M^2< 0$. The analytic continuation discussed in \cite{Dabholkar:1994ai} concerning $N$ gets translated to an analytic continuation in $\beta$. The arguments in favor of this continuation in \cite{Dabholkar:1994ai} are equivalent in our case to arguments of taking $\beta$ away from the orbifold values. In our language of field theory, the continuation in $\beta$ is quite natural. \\
We believe this is an important consistency check. To recapitulate, using the cigar WZW model and taking $k \to \infty$ we find the precise $\alpha'$ corrected action for the different types of strings. For the discrete orbifold temperatures however, we already know what the result should be \cite{Dabholkar:1994ai}\cite{Lowe:1994ah}. Since we are able to reproduce precisely the limiting free energy in the orbifold CFTs, we believe that indeed as argued above and in \cite{Giveon:2013ica}, the Rindler space thermal scalar action (including \emph{all} its $\alpha'$ corrections) can be obtained by taking $k \to \infty$ in the cigar CFT. The result is that type II superstrings do not receive $\alpha'$ corrections in their thermal scalar action, while bosonic strings do receive corrections. Without these, the orbifold result would not have been reproduced (the $1/N^2$ term in the exponent in (\ref{orbi}) would be missed). \\
Note also that we have gone full circle now: the $\tau_2 \to \infty$ limit should correspond to a state in the CFT that gives the dominant thermal behavior. From the field theory perspective we considered the thermal scalar action (with possible $\alpha'$ corrections) to give us this dominant contribution. But the link with the exact conformal description is lost. In this case we clearly see that these two descriptions match. \\
For heterotic strings, we also find precisely the same convergence criterion (\ref{convcrit}) as in the $\mathbb{C}/\mathbb{Z}_N$ orbifold models. Since the heterotic string thermal scalar action also includes a discrete momentum contribution, we again find it non-trivial to find a perfect match to the $\mathbb{C}/\mathbb{Z}_N$ orbifold. We believe that the heterotic thermal scalar action should also not receive $\alpha'$ corrections (like the type II superstring), though we lack a proof at the moment. The main complication is the subtleties with heterotic WZW models as discussed e.g. in \cite{Giveon:1993hm} and \cite{Sfetsos:1993bh}. \\
We note that, to treat $N\neq 1$, it suffices to simply take the correct value of $\beta = \frac{2\pi}{N}\sqrt{\alpha'}$; no extra (conical) corrections are present. If there would be conical corrections, we expect these to exist for these special cones as well. As discussed in section \ref{general}, the only place where $\beta$ enters the field theory action is in the $\partial_0$ derivatives and indeed we find here a $\beta^2$ contribution, which was the only type of $\beta$-dependent correction term we anticipated there. We believe this supports our expectation that we can safely take $\beta$ away from the CFT-values $\frac{2\pi}{N}\sqrt{\alpha'}$.\footnote{Let us elaborate on this point of view. Consider the field theory as discussed in general in section \ref{general}. Changing $\beta$ only affects the periodicity of one of the coordinates. For the T-dual background, the origin $\rho=0$ is not a fixpoint of the $U(1)$ rotation $\tau \to \tau + C$ with $C$ a real constant. So the T-dual geometry should not become `extra' singular just from the periodic identification, meaning that at $\rho=0$ we expect only the curvature singularity and this is not sensitive to the periodicity parameter $\beta$. More concretely, these arguments show that the only type of correction we can have in the thermal scalar action is 
\begin{equation}
\Delta S = f(\rho)\partial_\rho T \partial_\rho T^* + g(\rho)\beta^2 TT^{*} + h(\rho)TT^{*},
\end{equation}
with three unknown functions $f$, $g$ and $h$ that do not depend on $\beta$. Since we know the result at $\beta = \frac{2\pi}{N}\sqrt{\alpha'}$, we have $f=h=0$ and $g$ is the $\rho$-independent correction we found above. This holds now for \emph{all} values of $\beta$. In particular the $\alpha'$ corrections do not generate $\beta$-dependence, except the $\beta^2$ already discussed above. The apparent subtlety in these results is whether the thermal scalar action really can be determined only by the T-dual quantities, also off-shell. This seems to be the case and T-duality invariance is one of the biggest ideas used to construct off-shell descriptions of string theory (the so-called double field theory, see e.g. \cite{Aldazabal:2013sca} and references therein).} \\
We want to remark that there is a discrepancy in the wavefunctions: the orbifold twisted sector wavefunctions are localized at the tip of the cone, while in our case we find a string-scale spread around the tip when taking the $\beta$ continuation seriously.\footnote{Actually this might not be a real problem after all. Firstly, we notice that the link between these two wavefunctions is found by simply extracting the $\rho$-dependent part of the wavefunction. For instance for the bosonic (or type II) $w=\frac{1}{N}$ wavefunction $\psi_{n}$ and the twisted state wavefunction $F_n$ we would have (up to normalization): 
\begin{equation}
\psi_{n}(\tau,\rho,\bold{x}) = F_{n}(\tau,\bold{x})e^{-\frac{\rho^2}{4N}}L_{n}\left(\frac{\rho^2}{2N}\right).
\end{equation}
After this, we consider the resulting differential equation for $F_n(\tau,\bold{x})$, which has manifestly the same eigenvalue spectrum, as it should.\\
Secondly, our interest lies in the spread of the long string. This is written as a first-quantized path integral on the spatial submanifold. From this perspective, one could integrate out the $\rho$-field (schematically) as
\begin{equation}
\int \left[\mathcal{D}\tau\right]\left[\mathcal{D}\bold{x}\right]\left[\mathcal{D}\rho\right]e^{-S_p(\tau,\bold{x},\rho)} = \int \left[\mathcal{D}\tau\right]\left[\mathcal{D}\bold{x}\right]e^{-S_{eff}(\tau,\bold{x})}
\end{equation}
to obtain the twisted sector wavefunction point of view, though this is not what we want to do to distill the random walk picture.}
Nevertheless, we find it intriguing that we precisely reproduce the critical behavior of the partition function and we find a further explanation for the apparantly non-standard bosonic duality found in \cite{Dijkgraaf:1991ba}. 

\section{Unitarity constraints in Euclidean Rindler space}
\label{unitarity}
In this section we make the link between the $SL(2,\mathbb{R})/U(1)$ cigar and Euclidean Rindler space more precise in terms of conformal field theory language. In particular we want to analyze the `induced' unitarity constraints in Euclidean Rindler space. The cigar CFT states are characterized by several quantum numbers. The winding $w$ around the cigar and discrete momentum $n$ are collected in two linear combinations
\begin{equation}
m = \frac{n+kw}{2}, \quad \bar{m} = \frac{-n+kw}{2}, \quad n,w \in \mathbb{Z}.
\end{equation}
The quantum number $j$ is a measure for the radial momentum and is given as
\begin{align}
j &= -\frac{1}{2}+is, \quad s\in\mathbb{R} , \quad \text{continuous representations}, \\
j &= M - l , \quad l=1,2,...,\quad \text{discrete representations},
\end{align}
where $M = \text{min}(m,\bar{m})$ with $m,\bar{m} > 1/2$ \cite{Aharony:2004xn}.\footnote{\label{fn}There is also the option $M = \text{min}(\left|m\right|,\left|\bar{m}\right|)$ with $m,\bar{m} < -1/2$ \cite{Aharony:2004xn}, but we will not focus on this case since it will disappear when taking the flat $k\to\infty$ limit for the cases $w>0$. This sector is though relevant for the $w=-1$ state and we will comment on this further on. Note that only states with $\left|kw\right| > \left|n\right|$ are in the spectrum, these are the so-called winding dominated states \cite{Aharony:2004xn}.} We are interested in the discrete states on the cigar, so we consider the discrete representations. In this section we are interested in pure winding states, so $m = \bar{m} = M$. For the discrete representations $j$ has the following unitarity constraints:
\begin{align}
\label{unibos}
-\frac{1}{2} &< j < \frac{k-3}{2}, \quad \text{bosonic},\\
-\frac{1}{2} &< j < \frac{k-1}{2}, \quad \text{type II}.
\end{align}
The conformal weights of these states are given by
\begin{align}
h &= \frac{m^2}{k}-\frac{j(j+1)}{k-2}, \quad \bar{h} = \frac{\bar{m}^2}{k}-\frac{j(j+1)}{k-2} , \quad \text{bosonic},\\
h &= \frac{m^2}{k}-\frac{j(j+1)}{k}, \quad \bar{h} = \frac{\bar{m}^2}{k}-\frac{j(j+1)}{k} , \quad \text{type II}.
\end{align}

\subsection{Quantum numbers in Euclidean Rindler space}
Let us first analyze how these quantum numbers are reflected in the Rindler case.
We treat the type II superstring in this section.\footnote{The bosonic string case is analogous and we present the relevant formulas in appendix \ref{boswind}.} The eigenvalue equation we consider for $\beta = 2\pi\sqrt{\alpha'}$ is given by\footnote{We set $\alpha'=2$ in this section to conform to the conventions of \cite{Giveon:2013ica}. }
\begin{equation}
-\frac{\partial_\rho\left(\sinh\left(\sqrt{2/k}\rho\right)\partial_{\rho}T(\rho)\right)}{\sinh\left(\sqrt{2/k}\rho\right)} + \left(-1 + w^2\frac{k}{2}\tanh^2\left(\rho/\sqrt{2k}\right)\right)T(\rho)= \lambda T(\rho).
\end{equation}
The solution that does not blow up as $\rho \to \infty$ is given by
\begin{equation}
\label{expl}
T(\rho) \propto \frac{1}{\cosh\left(\frac{\rho}{\sqrt{2k}}\right)^{1+\sqrt{\omega}}}\,\, {\mbox{$_2$F$_1$}\left(\frac{\sqrt{\omega}+1+kw}{2},\frac{\sqrt{\omega}+1-kw}{2};\,\sqrt{\omega}+1;\,\frac{1}{\cosh\left(\frac{\rho}{\sqrt{2k}}\right)^{2}}\right)}.
\end{equation}
where $\omega = 1-2k-2k\lambda +k^2w^2$.
All normalizable states should behave near $\rho \to \infty$ as
\begin{equation}
\label{asymp}
\psi \propto \exp\left(-\sqrt{2/k}\left(j+1\right)\rho\right),
\end{equation}
since the background approaches a linear dilaton background there.
For a discrete pure winding state we have $j=\frac{kw}{2}-l$ where $l=1,2,3...$.
Since $\mbox{$_2$F$_1$}(a,b;c;0) = 1$, the prefactor determines the entire asymptotic behavior. Identifying the asymptotic behavior of (\ref{expl}) with the required asymptotics of (\ref{asymp}) gives
\begin{equation}
\label{bound}
\sqrt{\omega} = kw-2l+1,
\end{equation}
leading to 
\begin{equation}
\lambda = -\frac{2l(l-1)}{k} +2wl -w-1.
\end{equation}
For $k\to\infty$ the first term drops out and we are left with
\begin{equation}
\lambda \approx 2wl -w-1,\quad l=1,2\hdots 
\end{equation}
Setting $n=l-1$, we obtain
\begin{equation}
\lambda \approx 2wn +w-1,\quad n=0,1\hdots 
\end{equation}
which coincides with the discrete spectrum (\ref{superspectrum}) in Euclidean Rindler space. 
The condition to get the discrete states also implies that the second argument of the hypergeometric function becomes a negative (or zero) integer. This causes the hypergeometric function to become a polynomial and this is well-behaved also for $\rho \to 0$.\\
Note that there are only a finite number of discrete states since $l$ is bounded from above by the requirement that the r.h.s. of (\ref{bound}) should be positive. As $k$ increases, more values of $l$ are allowed and in the limit $k\to\infty$, $l$ becomes effectively unbounded. \\
The continuous states can be found for $\omega < 0$. This corresponds to a critical eigenvalue
\begin{equation}
\lambda^* = \frac{kw^2}{2} + \frac{1}{2k}-1,
\end{equation}
where $\lambda > \lambda^*$ corresponds to the continuous spectrum. Taking $k \to \infty$ gives $\lambda^* \to +\infty$ and the continuous states disappear.\\
It is instructive to see the above identification of $n$ and $l-1$ explicitly for the eigenfunctions. 
First consider the $w=1$, $l=1$ eigenfunction as given by \cite{Giveon:2013ica}:
\begin{equation}
T(\rho) \propto \frac{1}{\cosh\left(\frac{\rho}{\sqrt{2k}}\right)^k}.
\end{equation}
Taking $\rho \to \infty$ immediately gives
\begin{equation}
T(\rho) \propto \exp\left(-\sqrt{k/2}\rho\right)
\end{equation}
as it should be. Alternatively taking $k \to \infty$ gives
\begin{equation}
T(\rho) \propto \exp\left(-\rho^2/4\right).
\end{equation}
This identifies the $l=1$ wavefunction with the $n=0$ Rindler eigenfunction.\\
As another example, consider the $w=1$, $l=2$ bound state eigenfunction:
\begin{equation}
T(\rho) \propto \frac{(k-1) - (k-2)\cosh\left(\frac{\rho}{\sqrt{2k}}\right)^2}{\cosh\left(\frac{\rho}{\sqrt{2k}}\right)^k}.
\end{equation}
The $\rho \to \infty$ limit gives
\begin{equation}
T(\rho) \propto \exp\left(-\sqrt{2/k}\left(k/2-1\right)\rho\right).
\end{equation}
while taking $k \to \infty$ yields
\begin{equation}
T(\rho) \propto L_1\left(\rho^2/2\right)\exp\left(-\rho^2/4\right),
\end{equation}
which is precisely the $n=1$ eigenfunction in Rindler space. \\
We want to remark that the $w=-1$ state can be found on the cigar with the quantum numbers $j = M - l$ where this time $M = \left|-\frac{k}{2}\right|$ as we discussed in footnote \ref{fn} earlier. This yields $j = \frac{k}{2} - l$ and the asymptotic behavior is the same as that of the $w=1$ state. The Euclidean Rindler wavefunction is also the same as that of the $w=1$ state.

\subsection{Cigar spectrum in the $k\to\infty$ limit}
In this section, we take the large $k$ limit explicitly in the cigar spectrum. Our goal is to see the Rindler states explicitly appear and to further elaborate on the effect of the unitarity bounds.
For bosonic strings, we consider the cigar spectrum with some flat spectator dimensions to identify the tachyonic character of a state in the large $k$ limit. We have
\begin{equation}
h = -\frac{\alpha'M^2}{4} + \frac{m^2}{k}-\frac{j(j+1)}{k-2} =1.
\end{equation}
Setting $j=\frac{kw}{2}-l$, $l=1$ and $m=j+l$ we obtain
\begin{equation}
-\frac{\alpha'M^2}{4} + \frac{kw^2}{4}-\frac{(\frac{kw}{2}-1)(\frac{kw}{2})}{k-2} =1,
\end{equation}
and so (taking $k \to \infty$)
\begin{equation}
-\frac{\alpha'M^2}{4} + \frac{kw^2}{4}- \frac{kw^2}{4} - \frac{w^2}{2} + \frac{w}{2} =1
\end{equation}
or 
\begin{equation}
M^2 =  \frac{2}{\alpha'}\left( -w^2 + w - 2 \right).
\end{equation}
General $l$ would yield
\begin{equation}
M^2 =  \frac{2}{\alpha'}\left( -w^2 + (2l-1)w - 2 \right).
\end{equation}
For bosonic strings on the cigar, the $l=1$ sector is prohibited by the unitarity constraints (\ref{unibos}) in the $k\to\infty$ limit.
The bosonic unitarity constraint is $-1/2 < j < \frac{k-3}{2}$, where for winding states $j= \frac{kw}{2}-l$.
Setting $l=1$ and $w=1$ gives $j= \frac{k-2}{2}$ which violates the upper bound. Setting $l=2$ and $w=1$ on the other hand gives $j= \frac{k-4}{2}$ which is a valid string state. All higher values of $l$ are also allowed. Higher winding modes always violate the upper unitarity bound when taking $k\to \infty$.\\
Let us consider $\mathbb{Z}_N$ orbifolds of the $SL(2,\mathbb{R})/U(1)$ cigar. In this case, one introduces a conical singularity at the tip of the cigar. We expect that taking $k\to\infty$ in this CFT should coincide with the flat space $\mathbb{C}/\mathbb{Z}_N$ orbifolds.
To study orbifolds of the cigar, the effect of twisting the CFT is simply the change $w \to \frac{w}{N}$. That is, we should allow fractional winding numbers \cite{Martinec:2001cf}\cite{Son:2001qm}. States that have $w \notin N\mathbb{N}$ are the twisted sectors. The others are untwisted. We elaborate on this orbifolding procedure in appendix \ref{orbifoldpart} using the known cigar partition function \cite{Hanany:2002ev}. The cigar orbifold CFT in the $l=1$ sector thus has masses 
\begin{equation}
M^2 =  \frac{2}{\alpha'}\left( -\frac{w^2}{N^2} + \frac{w}{N} - 2 \right).
\end{equation}
In this case, $j= \frac{kw}{2N}-l$ and for $w=0\hdots N-1$, the $l=1$ sector satisfies the upper unitarity bound. These states can be identified with the twisted sector primaries of the $\mathbb{C}/\mathbb{Z}_N$ orbifold. The $w=N$ sector only starts with $l=2$ like in the unorbifolded case discussed above. Sectors with $w>N$ are to be excluded by the unitarity bound.\\
The analogous equations for type II superstrings are
\begin{equation}
-\frac{\alpha'M^2}{4} + \frac{m^2}{k}-\frac{j(j+1)}{k} =1/2.
\end{equation}
Again we set $j=\frac{kw}{2}-l$, $l=1$ and $m=j+l$ to obtain (taking $k \to \infty$):
\begin{equation}
-\frac{\alpha'M^2}{4} + \frac{kw^2}{4}- \frac{kw^2}{4} + \frac{w}{2} =1/2
\end{equation}
or 
\begin{equation}
M^2 =  \frac{2}{\alpha'}\left(  w - 1 \right).
\end{equation}
For general $l$ we would have
\begin{equation}
M^2 =  \frac{2}{\alpha'}\left( (2l-1)w - 1 \right).
\end{equation}
The unitarity constraint for type II superstring is $-1/2 < j < \frac{k-1}{2}$. In this case, $w=1$ and $l=1$ is in the spectrum (as are higher values of $l$). Higher winding modes on the other hand are again not allowed. We treat the discrete momentum modes and mixed momentum-winding modes in appendix \ref{spectr} to conclude our study of the NS-NS spectrum of conformal primaries in this space.\\
To summarize, all types of strings do not include higher winding modes $\left|w\right| > 1$ in the string spectrum. The $\left|w\right| = 1$ mode differs for bosonic strings and type II strings: in the bosonic case, $l=2,3,\hdots$ are allowed while in the type II case, $l=1,2,\hdots$ are allowed. This implies that in the bosonic Rindler eigenvalues we should drop the $n=0$ term and the $n=1$ contribution becomes the lowest eigenvalue. For flat $\mathbb{C}/\mathbb{Z}_N$ orbifolds, the twisted sectors $w=1\hdots N-1$ have $l=1,2,\hdots$ so these start with $n=0$ in the Rindler case. The $w=N$ sector only starts at $l=2$ for bosonic strings and $l=1$ for superstrings. The $w>N$ sectors are not present. 

\subsection{The bosonic Rindler string revisited}
\begin{figure}[h]
\centering
\includegraphics[width=13cm]{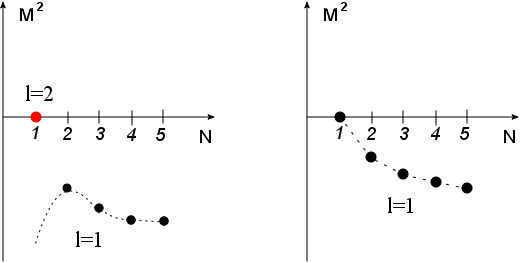}
\caption{Left figure: Most tachyonic state in the bosonic case versus the orbifold number $N$. The black dots are the $l=1$ modes. For $N=1$ the $l=1$ mode is absent and one should consider the $l=2$ mode (the red dot). The continuation in $N$ (the dashed line) suggests the entropy is divergent even though the free energy itself is not, since this only starts at $l=2$ for $N=1$. Right figure: Most tachyonic state in the type II superstring case. In this case, nothing special occurs and the $l=1$ mode is present both for the free energy and for the entropy.}
\label{cartoon}
\end{figure}

Let us reanalyze the critical behavior of the bosonic Rindler string, now incorporating these unitarity constraints. The lowest Rindler mode is the $n=1$ mode
\begin{equation}
\psi_{1}(\rho) \propto \exp\left(-\frac{\beta\rho^2}{4\pi\alpha'^{3/2}}\right)L_{1}\left(\frac{\beta\rho^2}{2\pi\alpha'^{3/2}}\right), \quad \lambda_{1} = \frac{3\beta - 4\pi\sqrt{\alpha'} - \frac{\beta^2}{2\pi\sqrt{\alpha'}}}{\pi \alpha'^{3/2}},
\end{equation}
where $L_1(x) = 1-x$ is the Laguerre polynomial of the first degree. This mode has a width
\begin{equation}
\left\langle \rho^2\right\rangle = \frac{6\pi\alpha'^{3/2}}{\beta}.
\end{equation}
The size of the wavefunction is a factor of $\sqrt{3}$ larger than the $n=0$ width. The eigenvalue is zero for $\beta = 2\pi\sqrt{\alpha'}$ and $\beta = 4\pi\sqrt{\alpha'}$. Values of $\beta$ in between these two values give a postive eigenvalue $\lambda_1$, while values of $\beta$ outside this range give rise to a negative eigenvalue and hence a divergence.
We conclude that for the canonical Rindler temperature, the free energy becomes marginally convergent, just like for the type II and heterotic strings.\\
However, as soon as one decreases $\beta$ (like for the orbifold CFTs), the $l=1$ mode reappears in the spectrum and the bosonic string free energy diverges. In particular, when computing the entropy as a derivative in $\beta$ of the free energy, we would expect a divergence. It is (presumably) only at precisely $\beta=\beta_R$ that the $l=1$ mode drops from the spectrum. For this reason, the bosonic thermal scalar is still effectively tachyonic. The situation is sketched in figure \ref{cartoon}.
Such restrictions in the Rindler quantum number $n$ are quite strange from a particle perspective. In particular the random walk picture has the caveat that we should drop the lowest mode of the associated wave operator. We see that string theory is more subtle than any limiting field theory might suggest.\\
Nonetheless, the analysis in section \ref{orbifold} concerning the $\mathbb{C}/\mathbb{Z}_N$ orbifolds remains valid (since the twisted sectors do have $l=1$ in the spectrum). The analysis concerning the thermal scalar action in section \ref{alphaprime} also remains valid: it is only the eigenmodes of this action that are subject to the unitarity constraints.\\
Perhaps even more importantly, the type II superstrings in Euclidean Rindler space do not suffer from this restriction on $l$ and ultimately we are interested in these. 

\section{Discussion}
\label{discussion}
We make several remarks concerning these results.\\

In appendix \ref{GM} we translate the previous results to the Schwarzschild normalization of Rindler space. We find that the state has a string width around the event horizon and the Hagedorn temperature for type II and heterotic superstrings becomes $\beta_{H} = 8\pi GM$ which is the Hawking temperature of the black hole. This temperature is much lower than the string scale unlike flat space Hagedorn temperatures. Physically, the Hawking temperature is the temperature set at $G_{00}=1$. So for Schwarzschild black holes, this temperature is the one at infinity. The size of the thermal circle goes all the way to zero at the horizon. So even choosing non-stringscale temperatures at infinity gives local stringscale temperatures near the tip of the cigar. Apparently this causes the singly wound state to be precisely massless at the Hawking temperature.\\

It is quite remarkable that for both type II superstrings and heterotic strings, the Rindler Hagedorn temperature equals the canonical Rindler temperature. It would have been surprising to obtain a divergence in these cases, since this would indicate an instability in thermodynamics and thus an instability in the theory itself. Quite reassuringly, there are no thermal tachyons in both these cases. No tachyon condensation process occurs, although an IR divergence can still occur depending on whether there are compact dimensions or not as we discussed above. Though it is peculiar that in both cases we obtain \emph{precisely} `marginal' stability. For both these string types, the winding string lives close to the event horizon. Since there is no tachyon anywhere in spacetime in these cases, this region does not correspond to a condensate but instead the marginal state is a bound state. Note that the localized state is in the perturbative spectrum of Euclidean Rindler space and its classical value is set to zero. \\ 

If we look back at the operator (\ref{opera}), we notice that the reason the thermal scalar is localized near the horizon is the $\rho^2$ potential and this is present since we study $w=\pm1$. The `normal' particles (having $w=n=0$) do not have this potential and can freely propagate outward (if they overcome the centrifugal barrier as discussed in e.g. \cite{Susskind:2005js}). Besides these, there are also $n\neq0$ states that propagate to infinity. For the type II string, the wavefunctions of these states and their relation to the cigar CFT can be found in appendix \ref{spectr}. This is a very important difference between these fields: non-winding states can propagate to infinity whereas the thermal scalar on the other hand is bound to the horizon. It is to be interpreted in the Lorentzian signature as a long highly excited string located at a string length from the horizon. Considering the thermal zone of a black hole, the thermal scalar is living much closer to the horizon, effectively outside the reach of low energy quantum field theory. This makes it a natural microscopic candidate for the stretched horizon (or the membrane). Note that classically, the black hole membrane is located at an arbitrary radial location that is sufficiently small \cite{Thorne:1986iy}. In our case, the stretched horizon has a fixed location at a string length outside the horizon. \\

The ground state wavefunctions determined above have a width equal to (or of the order of) $\sqrt{\alpha'}$ for $\beta = 2 \pi \sqrt{\alpha'}$, so consistency with the Rindler approximation requires this to be smaller than $4GM$. We thus arrive at the following equivalent consistency requirements 
\begin{equation}
\alpha' \ll G^2M^2 \quad \Leftrightarrow \quad l_{s} \ll R_{H} \quad \Leftrightarrow \quad T_{Hawking} \ll T_{flat space Hagedorn}.
\end{equation}
So if the asymptotic temperature of the black hole is much smaller than the \emph{flat space} Hagedorn temperature, we are safe to use Rindler space. This is equivalent to choosing a black hole much larger than the string length. We conclude that our picture is only valid when the horizon size is much larger than the width of the dominant string state. 
The above was applied to a Schwarzschild black hole but it applies equally well to (large) AdS black holes. We are more interested in the latter since these are thermodynamically stable and we can use the $AdS$ spacetime as a `container' to mitigate the ever-present Jeans instability \cite{Barbon:2001di}\cite{Barbon:2002nw}. \\
We also point out that when we consider Rindler spacetime, we have no way of seeing the ground state width diverge as was argued in \cite{Kutasov:2005rr} to happen for sufficiently small black holes. This is obviously caused by the fact that Rindler spacetime has an ever increasing circumference of the thermal circle, while for a real black hole the circumference asymptotes to the radius corresponding to the Hawking temperature. The authors of \cite{Kutasov:2005rr} and \cite{Giveon:2005jv} interpret this diverging thermal scalar width as the black hole-string correspondence point. So we do not have the chance to study the black hole evaporation process and the correspondence principle as put forth by \cite{Horowitz:1996nw}. \\

From the Rindler example, we can learn some lessons regarding the higher winding modes. Naively, when considering a conical space (or its smooth cigar-like cousin), one can imagine that \emph{all} winding modes are tachyonic because the radius of the thermal circle shrinks all the way to zero. Thus naively we cross all Hagedorn transition temperatures for all winding modes. This reasoning is wrong. 
The $SL(2,\mathbb{R})/U(1)$ cigar CFT spectrum explicitly shows that this is not the case: it depends on the quantum numbers considered. When taking the flat limit $k\to\infty$, we instead see that all higher winding modes are simply absent from the string spectrum. This is a feature to which the field theory action is a priori insensitive. The singly wound mode is a marginal state and hence is not tachyonic. \\

This entire discussion has been for strings with all spectator dimensions geometrically flat. One could ask whether the above picture is altered when compactifying on geometrically non-trivial spaces. Let us split spacetime in a four dimensional (for string phenomenology) or five dimensional (for holographic purposes) spacetime and a compact internal manifold. 
Smooth compactifications (e.g. on a compact Calabi-Yau manifold $\mathcal{M}$) that are compact unitary CFTs do not alter our conclusions. The reason is that the conformal weights are $h \geq 0$ in this case and $h=0$ is in the spectrum (this is the unit operator and this state is automatically normalizable due to the compactness). Due to the compactness, we also do not introduce continuous quantum numbers. Thus the convergence calculation remains the same and the winding mode retains its character (marginally convergent for all string types). More explicitly, consider the partition function
\begin{equation}
Z = \int_{F}\frac{d\tau_1d\tau_2}{2\tau_2}Z_{matter}(\tau)Z_{gh}(\tau)Z_{compact}(\tau),
\end{equation}
where
\begin{align}
Z_{matter}(\tau) &= \text{Tr}\left(q^{L_0 -c_m/24}\bar{q}^{\bar{L}_0-\bar{c}_m/24}\right), \\
Z_{gh}(\tau) &= \text{Tr}\left(q^{L_0 -c_g/24}\bar{q}^{\bar{L}_0-\bar{c}_g/24}\right),\\
Z_{compact}(\tau) &= \text{Tr}\left(q^{L_0 -c_{comp}/24}\bar{q}^{\bar{L}_0-\bar{c}_{comp}/24}\right),
\end{align}
with $c_m + c_{g} + c_{comp} = 0$. We immediately cancel the $c$-dependent factors in the trace and in the limit $\tau_2 \to \infty$, upon dropping the $c$-dependent factors, we have that
\begin{equation}
\tilde{Z}_{compact}(\tau) = \text{Tr}\left(q^{L_0}\bar{q}^{\bar{L}_0}\right) \to 1.
\end{equation}
In this limit the compact part drops out of the partition function. This has an analogous manifestation in the heat kernel picture. Since the heat kernel factorizes, the ground state wavefunction becomes a product of the two ground states and the shape of the tachyon wavefunction in the $\rho$ direction remains the same. For the compact dimensions we have
\begin{equation}
K(\tau_2) = \int dV K(x,x;\tau_2) = \sum_n\int dV \psi_n(x)\psi_n(x)^{*}e^{-E_{n}\tau_2} \to 1.
\end{equation}
when we assume that we do not have negative eigenvalues of the operator that gives us this heat kernel. The constant function is an eigenfunction with eigenvalue zero, so its wavefunction is constant and equal to $\psi_0 = \frac{1}{\sqrt{V}}$ (up to a phase). The critical behavior and the Hagedorn temperature are indeed not modified and the ground state wavefunction is uniformly distributed over the compact manifold. The random walk behavior is entirely determined by the non-compact part.\\
As an explicit example, consider a $SU(2)_k$ WZW model as (part of) the compact CFT. The primaries are discrete and have conformal weights
\begin{align}
h &= \frac{1}{k+2} j(j+1), \quad \bar{h} = \frac{1}{k+2}\bar{j}(\bar{j}+1), \quad \text{bosonic string}, \\
h &= \frac{1}{k} j(j+1), \quad \bar{h} = \frac{1}{k}\bar{j}(\bar{j}+1), \quad \text{type II superstring},
\end{align}
where $0\leq j,\bar{j}\leq k/2$. Which primaries appear in the string spectrum is irrelevant for our discussion (this corresponds to choosing a specific modular invariant). We clearly see that the minimal conformal weight is indeed zero, as it should for a unitary compact CFT. \\
From the heat kernel point of view, the scalar Laplacian on $S_3$ has eigenvalues $-l(l+2)$ where $0 \leq l$ an integer. Clearly the lowest eigenvalue is zero and the Hagedorn temperature remains unchanged. \\

Now that the dust has settled, we make a comparison between the different approaches to Rindler near-Hagedorn thermodynamics. \\
As a first approach we have the random walk picture as made explicit by \cite{Kruczenski:2005pj}\cite{theory} and reviewed in section \ref{pathderiv}. This realizes the picture as proposed by Susskind \cite{Susskind:2005js} that highly excited strings can be described as random walks and that near the Hagedorn transition, the string gas recombines itself into a single highly excited string.\footnote{Again with the caveats we discussed before.} The random walk of the thermal scalar in the Euclidean picture traces out the spatial form of the long string in the Lorentzian picture. The $\tau_2 \to \infty$ limit corresponds to a long walk, and thus to a long string in the Lorentzian picture. The subtle point is however that the fate of the correction terms of the worldsheet action is unclear. Also, since we started with the gauge-fixed string path integral, we cannot go off-shell.\\
The other approach is to start with the field theory action, compute the one-loop amplitude and convert this to a first-quantized form. When properly taking into account the $G_{00}$ metric component and $\alpha'$ corrections, we get a modified random walk picture and we interpret the modifications as the correction terms from above. So from this approach, we do get information about the corrections. Also, since this is a field theory, we appear to have no problem in going off-shell. However, here we have trouble in interpreting the required Wick rotation to return to the Lorentzian picture as discussed in \cite{Kutasov:2000jp}\cite{Giveon:2012kp}.\\
We see that the delicate points of one approach are explained by the other approach. So combining these two viewpoints, we have all ingredients to fully realize the Rindler string as a (modified) random walker that is confined to a string length from the event horizon. Thus this realizes (at genus one) the picture put forward by Susskind.

\section{Open problems and speculations}
\label{open}
We take a look into some puzzles and unsolved problems that we face.

\begin{itemize}

\item{It would be nice to further clarify the precise correspondence between the cigar CFT and the $\mathbb{C}/\mathbb{Z}_N$ orbifold. A better understanding of the heterotic case and why it also matches to the flat orbifold without any $\alpha'$ corrections would be helpful. A study of this is postponed to future work.}

\item{
The authors of \cite{Kutasov:2000jp} argued that on the $SL(2,\mathbb{R})/U(1)$ cigar in the type II superstring case, the marginal state should become tachyonic when considering higher loop effects: the tree level mass should get corrections that drive it tachyonic. This idea arose in order to agree with (and explain) the unstable little string thermodynamics. The one-loop free energy is \emph{only} sensitive to the tree-level spectrum, so indeed this is not divergent. The higher genus contributions to the free energy however, are expected to diverge. Whether this is the case or not is, as far as we know, not explicitly known. Note that an alternative explanation of the unstable little string thermodynamics was given in \cite{Rangamani:2001ir}\cite{Buchel:2001dg} where the instability was attributed to a gravitational zero-mode. This solution would not require the winding string zero-mode to become tachyonic at higher genera. 
}

\item{In some less recent literature (see e.g. \cite{Parentani:1989gq}, \cite{Emparan:1994bt} and \cite{McGuigan:1994tg}), it was found that the Green propagator (and the energy-momentum tensor) should diverge for the canonical Rindler temperature. Their arguments do not care about the specifics of the bosonic string theory and are equally valid for superstrings and heterotic strings. So in these papers, it is suggested that the region near the horizon causes divergences in several quantities, corresponding to a \emph{maximal acceleration} of strings near black hole horizons. In \cite{McGuigan:1994tg} it was argued that this provides a natural cut-off to the theory. It would be very interesting to know the link to our work.}

\item{One could think about expanding the random walk path integral picture to include Ramond-Ramond backgrounds. The goal would be to again identify the membrane with the long string(s). Besides the technical difficulties this would entail, we would like to point out a conceptual problem with this. For a neutral black hole, the above arguments suggest that indeed a long string dominates the thermal ensemble, exactly like in flat spacetime. This agrees with the black hole correspondence principle \cite{Horowitz:1996nw} where a neutral black hole evaporates into a long highly excited string.\footnote{In \cite{Susskind:2005js} this is viewed as the same string that constitutes the stretched membrane.} When we consider RR charged black holes however, the correspondence principle suggests a match to a gas of open strings living on a D-brane. This suggests that we should augment the closed string random walk picture with open string sectors to find the stretched membrane for such black holes. The thermal scalar appears not to be sufficient to constitute the membrane in this case. 
This behavior is not universal for all charged black holes since for electrically NS-NS charged black holes, the correspondence principle \emph{does} suggest a match to long closed strings. \\
We finally remark that our study was also limited to non-extremal black holes which have Rindler space near their horizon. Extremal black holes typically develop an infinite throat (e.g. the extremal 4d Reissner-Nordstr\"om black hole has $AdS_2 \times S^2$ near-horizon geometry).
}

\end{itemize}

\section{Conclusion}
\label{conclusion}
In \cite{theory}, we reviewed and extended the path integral derivation of the random walk behavior of near-Hagedorn thermodynamics \cite{Kruczenski:2005pj}. Several questions arose in the process: is there really a winding mode in the string spectrum, especially if the space does not topologically support winding modes? Can we get a handle on the higher $\alpha'$-correction terms? In this paper, we found answers to these questions by analyzing Rindler spacetime as the near-horizon approximation of black holes. 

We found that the random walk gets corrected but we know precisely what the correction is (due to its relation to the $SL(2,\mathbb{R})/U(1)$ cigar). Moreover the resulting particle path integral can be exactly solved and we checked that it gives the same answer as the zero-mode of the Hamiltonian operator associated with it, which was recently found in a slightly different context in \cite{Giveon:2012kp}\cite{Giveon:2013ica}. We used one-loop convergence as a criterion to obtain the Hagedorn temperature and found the critical temperatures and critical behavior for the Rindler string gas of all string types (bosonic, type II and heterotic). All string types show critical Hagedorn behavior at the canonical Rindler temperature.\footnote{Keeping in mind the complications for the bosonic string.} We interpreted the stretched horizon as a Hagedorn string at string length from the event horizon whose temperature is redshifted to the Hawking temperature at infinity. To further substantiate the claims in \cite{Giveon:2013ica} regarding the $\alpha'$ corrections, we have shown that the corrections to the bosonic string thermal scalar action are precisely such that they reproduce correctly the flat $\mathbb{C}/\mathbb{Z}_N$ limit. Disregarding the bosonic corrections would yield a different answer. The unitarity constraints from the cigar get transferred to Euclidean Rindler space, where they in particular forbid any higher winding modes. \\

Our attitude towards this approach to string thermodynamics is the following. The method reduces string theory to a particle theory of the thermal scalar. This inherently brings back the UV divergence one typically encounters when evaluating thermodynamical quantities in black hole spacetimes \cite{'tHooft:1984re}\cite{Barbon:1994ej}\cite{Emparan:1994qa}. So it is not obvious how to work with the obtained expressions to get quantitative predictions on the free energy and entropy. What one pays for in this aspect, we gain in another: namely because we have a field theory, we are free to go off-shell. We obtain an off-shell description of the dominant behavior of string thermodynamics near black holes where we are free to vary $\beta$ \emph{without} having to worry about violating (super)conformal invariance. We can see that black holes when equilibrated with a gas of strings are surrounded by a thin stringsize shell of the thermal scalar, that we identify with the black hole membrane. This membrane differs for the different types of strings: for bosonic strings, we have the complications we addressed in section \ref{unitarity}: the lowest mode is not in the string spectrum and the free energy does not have a tachyonic divergence. Flat orbifolds on the other hand always include tachyonic divergences and this suggests quantities like the entropy also are tachyonically divergent. For both superstrings and heterotic strings, the situation is simpler and we have a marginal convergence: the membrane is composed of a zero-mode. The heterotic membrane is also slightly larger than the superstring membrane.\\

We finally compared the two approaches to Rindler thermodynamics (string path integral Fourier expansion versus field theory) and noted that, when combined, they describe the picture by Susskind of a single long string surrounding the event horizon, at least for the genus one worldsheet.

\section*{Acknowledgements}
The authors would like to thank Ben Craps, David Dudal and Lihui Liu for several valuable discussions and Amit Giveon and Nissan Itzhaki for useful correspondence. We also thank David Dudal for a careful reading of the manuscript. TM thanks the UGent Special Research Fund for financial support. The work of VIZ was partially supported by the RFBR grant 14-02-01185.

\appendix

\section{Schwarzschild normalization of Rindler space}
\label{GM}
We translate the results from sections \ref{typeII}, \ref{heterotic} and \ref{bosonic} to the Schwarzschild-normalized Rindler space:
\begin{equation}
ds^2 = -\left(\frac{\rho^2}{(4GM)^2}\right)dt^2 + d\rho^2 + d\mathbf{x}^2_{\perp}.
\end{equation}
For type II superstrings, the eigenfunctions and eigenvalues in these coordinates are
\begin{equation}
\psi_{n}(\rho) \propto \exp\left(-\frac{\beta\rho^2}{4\pi\alpha'(4GM)}\right)L_{n}\left(\frac{\beta\rho^2}{2\pi\alpha'(4GM)}\right), \quad \lambda_{n} = \frac{\beta - 2\pi(4GM)+2\beta n}{\pi \alpha'(4GM)}.
\end{equation}
The lowest eigenmode has the form
\begin{equation}
\psi_0 \propto \exp\left(-\frac{\beta\rho^2}{16\pi\alpha'GM}\right).
\end{equation}
For $\beta = \beta_R = 8\pi GM$, we find
\begin{equation}
\psi_0 \propto \exp\left(-\frac{\rho^2}{2\alpha'}\right)
\end{equation}
and we clearly see that the mode has stringscale width. \\
For heterotic strings, the eigenfunctions and eigenvalues become
\begin{align}
\psi_{n}(\rho) &\propto \rho^{\frac{\pi(4GM)}{\beta}}\exp\left(-\frac{\beta\rho^2}{4\pi\alpha'(4GM)}\right)L_{n}^{(a)}\left(\frac{\beta\rho^2}{2\pi\alpha'(4GM)}\right), \\
\quad \lambda_{n} &= \frac{\beta - 2\pi(4GM)+2\beta n}{\pi \alpha'(4GM)},
\end{align}
where $a=\frac{\pi(4GM)}{\beta}$.\\
Both for type II superstrings and heterotic strings, the Hagedorn temperature is equal to the canonical Rindler temperature which is (with this normalization) also equal to the Hawking temperature:
\begin{equation}
\beta_{Hawking} = 8\pi GM.
\end{equation}
For bosonic strings, the eigenfunctions are the same as those of the type II superstring, and the eigenvalues are given by
\begin{equation}
\lambda_{n} = \frac{\beta +2\beta n}{\pi \alpha'(4GM)} - \frac{4}{\alpha'},
\end{equation}
whereas the $\alpha'$-corrected eigenvalues are
\begin{equation}
\lambda_{n} = \frac{\beta  +2\beta n}{\pi \alpha'(4GM)} - \frac{4}{\alpha'} - \frac{\beta^2}{2\pi^2\alpha'^2}.
\end{equation}

\section{Explicit solution of the Rindler random walk path integral}
\label{appB}
In this appendix we explicity solve the random walk particle path integrals and show that they match with the field theory computations as they should. We start with bosonic strings. The thermal scalar action in Rindler space is governed by the particle action\footnote{Substitute $t \to t/(2\pi\alpha')$ in expression (\ref{act}).}
\begin{equation}
\label{pppi}
S = \frac{1}{2}\int_{0}^{T}dt\left[\dot{\rho}^2+\frac{\beta^2\rho^2}{4\pi^2\alpha'^3}-\frac{1}{4\rho^2}-\frac{4}{\alpha'}- 2\frac{\beta^2}{4\pi^2\alpha'^2}\right].
\end{equation}
This particle path integral lives on a half-line in a harmonic oscillator and in a $1/\rho^2$ potential. This model can be exactly solved \cite{Khandekar:1975}. The non-trivial part of the particle action is given by
\begin{equation}
S = \int_{0}^{T}dt\left(\frac{1}{2}m\dot{\rho}^2 + \frac{1}{2}m\omega^2\rho^2 + \frac{g}{\rho^2}\right).
\end{equation}
The resulting heat kernel is given by
\begin{equation}
K(\rho,T|\rho,0) = \rho\left[\frac{m\omega}{\sinh(\omega T)}\right]\exp\left(-m\omega \rho^2\coth(\omega T)\right)I_{\kappa}\left(\frac{\rho^2m\omega}{\sinh(\omega T)}\right)
\end{equation}
where
\begin{equation}
m=1,\quad \omega^2 = \frac{\beta^2}{4\pi^2\alpha'^3},\quad g= -\frac{1}{8}, \quad \kappa^2 = 2mg + 1/4 = 0.
\end{equation}
Since $g < 0$ the inverse potential is attractive. Note that since $\kappa = 0$, we have precisely saturated the stability limit of this problem: any stronger attraction would result in the `fall to the center' as discribed in \cite{Khandekar:1975}.
This propagator (heat kernel) was derived in \cite{Khandekar:1975} using an explicit path integration. It was noted there that due to the singular perturbation $\propto \frac{1}{\rho^2}$, there is no `contact' between the states living at $\rho>0$ and those at $\rho<0$. The integration range of the intermediate integrations in the path integral is from $\rho=0$ to $\rho \to \infty$. So this explicitly demonstrates that our space is restricted to a half-line. If we neglect the singular perturbation, the origin $\rho=0$ would be regular, and we would need to think about which boundary conditions to impose at the origin. In our case here, the nature of the potential solves this problem. Notice that this singular perturbation term is precisely the correction term $K(x)$ that we discussed previously, so discarding it would fundamentally alter the problem.\\
One easily checks that the integrals converge for $\rho \to 0$ and for $\rho \to \infty$.\footnote{The modified Bessel function has the following behavior
\begin{align}
I_{\alpha}(x) &= \frac{1}{\Gamma(\alpha+1)}\left(\frac{x}{2}\right)^{\alpha},\quad x \approx 0, \\
I_{\alpha}(x) &= \frac{e^{x}}{\sqrt{2\pi x}},\quad x \gg 1.
\end{align}} The $T \to 0$ limit diverges, corresponding to the field theory UV divergence.
The integral over $\rho$ can be done and results in\footnote{$\int_{0}^{+\infty}dx \exp\left(-\omega x \coth(\omega T)\right)I_{A}\left[\omega x / \sinh(\omega T)\right] = \frac{1}{\omega}\exp(-A\omega T)$.} 
\begin{equation}
\int_{0}^{+\infty}d\rho K(\rho,T|\rho,0) = \frac{1}{2\omega}\frac{\omega}{\sinh(\omega T)}.
\end{equation}
When examining the interesting $T \to \infty$ limit, the traced heat kernel becomes\footnote{Here we reincluded the $\rho$-independent contributions to the action (\ref{pppi}).}
\begin{equation}
\sim e^{-\omega T}e^{\frac{2}{\alpha'}T}e^{\frac{\beta^2}{4\pi^2\alpha'^2}T}.
\end{equation}
which indeed is the same as equation (\ref{meth1alpha}) as soon as one substitutes $T \to T/2$ in the $T$-integral in (\ref{freebosonic}).\\
Also the Schr\"odinger equation corresponding to the particle path integral (\ref{pppi}) can be directly solved \cite{Khandekar:1975} and yields the following eigenfunctions and eigenvalues:
\begin{equation}
\phi_{n}(\rho) \propto \rho^{\kappa + 1/2}L^{(\kappa)}_{n}(m\omega \rho^2)e^{-\frac{m\omega}{2}\rho^2}, \quad \lambda_{n} = (2n+1+\kappa)\omega
\end{equation}
where $n=0,1,2,\hdots$ We observe that these eigenfunctions differ from those in equation (\ref{eigenf}) by a factor of $G_{00}^{1/4} \propto \sqrt{\rho}$ which is a consequence of the different inner product on the Hilbert space in extracting the $\sqrt{G_{00}}$ from the measure in the field theory action as we discussed in \cite{theory}.\\
For type II superstrings, the changes are trivial: one only needs to substitute $\frac{4}{\alpha'} \to \frac{2}{\alpha'}$ for the covering space mass term in the particle path integral.\\
For heterotic strings, we can also compute the particle path integral. Starting with the particle path integral (\ref{pppi}), we again first perform the substitution $t \to t/(2\pi\alpha')$.
The extra discrete momentum term corresponds to a shift in the parameter $g$ of the particle path integral model
\begin{equation}
g \to g + \frac{\pi^2\alpha'}{2\beta^2}
\end{equation}
and this leads to
\begin{equation}
\kappa^2 = \frac{\pi^2\alpha'}{\beta^2}.
\end{equation}
One readily finds that the resulting traced heat kernel and eigenvectors and eigenvalues of the associated Schr\"odinger operator are consistent with those found in section \ref{heterotic}.

\section{Thermal scalar dominance}
\label{dom}
We will clarify in what sense the thermal scalar dominates the free energy in the canonical ensemble. The main question one should ask is whether the target space is compact or not. This distinction can be understood on quite general terms from a random walk perspective \cite{Barbon:2004dd} (see also \cite{Deo:1991mp} for more extensive discussions). In this section we illustrate these claims with the simple flat space example and we explicitly demonstrate the significance of the compactness of the target space. For simplicity we only consider bosonic strings.\\
All divergences in the free energy expression come from the $\tau_2 \to \infty$ limit in the fundamental modular domain.
A flat non-compact dimension gives a factor
\begin{equation}
Z_X(\tau) = \frac{\left|\eta\right|^{-2}}{\sqrt{4\pi^2\alpha'\tau_2}} \to \frac{e^{\pi\tau_2/6}}{\sqrt{4\pi^2\alpha'\tau_2}},
\end{equation}
while a flat compact dimension gives
\begin{equation}
2\pi RZ_X(\tau) \sum_{n,m}\exp\left(-\pi R^2\frac{\left|m-n\tau\right|^2}{\alpha'\tau_2}\right) \to e^{\pi\tau_2/6},
\end{equation}
where in both cases we have taken the $\tau_2 \to \infty$ limit.
For the temperature-dependent part we have dropped the $n=0$ contribution.
If all dimensions are non-compact, the free energy in flat space can be written as
\begin{equation}
F = -V\int_{F}\frac{d\tau d\bar{\tau}}{2\tau_2}\frac{1}{(4\pi^2\alpha'\tau_2)^{25/2}}\left|\eta(\tau)\right|^{-46}\frac{1}{(4\pi^2\alpha'\tau_2)^{1/2}}\left|\eta(\tau)\right|^{-2}\sum_{n,m}\exp\left(-\beta^2\frac{\left|m-n\tau\right|^2}{4\pi\alpha'\tau_2}\right)
\end{equation}
and yields in the critical limit
\begin{equation}
F \to -V\int_{F}\frac{d\tau d\bar{\tau}}{\tau_2}\frac{1}{(4\pi^2\alpha'\tau_2)^{25/2}}e^{4\pi\tau_2}e^{-\frac{\beta^2\left|\tau\right|^2}{4\pi\alpha'\tau_2}}.
\end{equation}
We see that this expression converges at $\beta = \beta_H$ due to the suppressing $\tau_2$ factors from the non-compact dimensions.
If on the other hand we choose only compact dimensions, we get
\begin{equation}
\label{comp}
F = -V\int_{F}\frac{d\tau d\bar{\tau}}{2\tau_2}\frac{1}{(4\pi^2\alpha'\tau_2)^{13}}\left|\eta(\tau)\right|^{-48}\prod_{i}\sum_{n_i,m_i}\exp\left(-\pi R_i^2\frac{\left|m_i-n_i\tau\right|^2}{\alpha'\tau_2}\right),
\end{equation}
and this gives
\begin{equation}
F \to -V\int_{F}\frac{d\tau d\bar{\tau}}{\tau_2}e^{4\pi\tau_2}e^{-\frac{\beta^2\left|\tau\right|^2}{4\pi\alpha'\tau_2}}.
\end{equation}
For $\tau_2 \to \infty$, this diverges logarithmically at $\beta = \beta_H$. Choosing at least one non-compact dimension gives convergence. We also remark that the two limits $\tau_2 \to \infty$ and $R_{i} \to \infty$ do not commute. \\
Note that the `normal' massless non-winding strings contribute to the free energy (\ref{comp}) as
\begin{equation}
F \to -V\int_{F}\frac{d\tau d\bar{\tau}}{\tau_2}.
\end{equation}
This also diverges logarithmically, but gives a temperature-independent contribution to the free energy and we are not interested in this.\footnote{These drop out anyway when computing the entropy.}
So for fully compact spaces, when the temperature is close to the Hagedorn temperature (but still smaller), the free energy contribution from the thermal scalar dominates the full free energy as
\begin{equation}
F \propto -\ln(\beta-\beta_H) + \text{infinite but independent of $\beta$} + \text{finite}.
\end{equation}
More generally, for fully compact spaces (not necessarily flat) we expect a behavior $F \propto -\ln(\beta - \beta_H)$ that diverges at $\beta = \beta_{H}$. Noncompact spaces do not yield such a divergence.\\
If at least one dimension is non-compact the free energy does not diverge and the thermal scalar does not dominate the free energy (since it remains finite): it gives the leading non-analytic behavior of the thermodynamic quantities. If on the other hand, all dimensions are compact the thermal scalar really takes over the entire thermodynamics. \\
In order to circumvent the Jeans instability, which is always present for sufficiently large thermal systems if we include gravity, we should consider only compact dimensions and the thermal scalar really represents the dominant contribution.

\section{Cigar orbifold partition function}
\label{orbifoldpart}
In this section we compute the partition function of $\mathbb{Z}_N$ orbifolds of the $SL(2,\mathbb{R})/U(1)$ model. We follow the computation of \cite{Hanany:2002ev} and indicate where differences occur. In \cite{Son:2001qm} the orbifolds of $AdS_3$ were considered which are closely related to the ones we study in this section. We are interested in these cigar orbifolds since we want to make sure that indeed orbifolding corresponds to including fractional winding numbers with the \emph{same} unitarity bounds as the ones obtained for the unorbifolded case. Also, we want to investigate whether some extra states occur like those in the (Lorentzian signature) $AdS_3$ orbifold model found in \cite{Son:2001qm}. For discussions concerning the superstring case, we refer to \cite{Eguchi:2010cb}. \\
The strategy is to write down an expression for the partition function by explicitly integrating the string path integral. Then this expression needs to be rewritten in terms of the $\widehat{SL(2,\mathbb{R})}$ characters such that it has the form 
\begin{equation}
Z(\tau) = \sum_i N_{i\tilde{i}}\chi_{i}(\tau) \chi_{\tilde{i}}^{*}(\tau)
\end{equation}
and we can identify the different states that occur in the string spectrum. Following \cite{Hanany:2002ev}, we start with a gauged WZW model with coordinates $\theta$, $\tilde{\theta}$ and $r$, where $\theta = \frac{1}{2}(\theta_L - \theta_R)$ and $\tilde{\theta} = \frac{1}{2}(\theta_L + \theta_R)$.
The gauge field degrees of freedom are translated into scalars $\rho$ and $\tilde{\rho}$ and we define two new coordinates
$\kappa = \theta + \rho$ and $\tilde{\kappa} = \tilde{\theta} - \tilde{\rho}$.
Next we do a coordinate transformation
\begin{align}
v &= \sinh(r/2) e^{i\kappa}, \\
\bar{v} &= \sinh(r/2) e^{-i\kappa}, \\
\phi &= i\tilde{\kappa} - \log\cosh(r/2).
\end{align}
The $\theta$ coordinate is identified with period $2\pi$ and represents the angular coordinate on the cigar. This periodicity is passed over to $\kappa$. If we want to consider $\mathbb{Z}_N$ orbifolds of the cigar, we should change the periodicity to $2\pi/N$. This means that we consider the sectors
\begin{equation}
v(z+2\pi) = H(v)(z)e^{i2\pi a/N}, \quad v(z+2\pi \tau) = H(v)(z)e^{i2\pi b/N},
\end{equation}
where $H(v)$ denotes the `normal' effect of the gauge holonomies on $v$ and $a,b = 0...N-1$ label the different sectors. Extracting the non-periodic parts from $v$, we obtain
\small
\begin{align}
v(z) &= \hat{v}\exp\left[-\frac{1}{4\tau_2}((u_1 \bar{\tau}-u_2)z - (u_1 \tau -u_2) \bar{z}) - \frac{1}{2\tau_2}\left(\left(\frac{a}{N}\bar{\tau}-\frac{b}{N}\right)z - \left(\frac{a}{N} \tau - \frac{b}{N}\right) \bar{z}\right)\right], \\
\bar{v}(z) &= \hat{\bar{v}}\exp\left[\frac{1}{4\tau_2}((u_1 \bar{\tau}-u_2)z - (u_1 \tau -u_2) \bar{z}) + \frac{1}{2\tau_2}\left(\left(\frac{a}{N}\bar{\tau}-\frac{b}{N}\right)z - \left(\frac{a}{N} \tau - \frac{b}{N}\right) \bar{z}\right)\right], \\
\rho(z) &= \hat{\rho} + \frac{1}{4\tau_2}((u_1 \bar{\tau}-u_2)z + (u_1 \tau -u_2) \bar{z}), \\
\phi(z) &= \hat{\phi} + \frac{1}{4\tau_2}((u_1 \bar{\tau}-u_2)z + (u_1 \tau -u_2) \bar{z}).
\end{align}
\normalsize
where $\hat{v}, \hat{\bar{v}}, \hat{\phi}, \hat{\rho}$ denote periodic fields on the torus.\\
When including all the contributions to the path integral \cite{Hanany:2002ev}, we arrive at
\begin{align}
Z &= \frac{1}{N}\sum_{a,b=0}^{N-1}2\sqrt{k(k-2)}\int_{F}\frac{d\tau d\bar{\tau}}{\tau_2} \int_{-\infty}^{+\infty}du_1du_2 \nonumber \\
&\sum_i q^{h_i}\bar{q}^{\bar{h}_i}e^{4\pi\tau_2(1-\frac{1}{4(k-2)}) -\frac{k\pi}{\tau_2}\left|u_1\tau -u_2\right|^2+2\pi\tau_2\tilde{u}_1^2} \nonumber \\
&\frac{1}{\left|\sin(\pi(\tilde{u}_1\tau - \tilde{u}_2))\right|^2}\left|\prod_{r=1}^{+\infty}\frac{(1-e^{2\pi i r \tau})^2}{(1-e^{2\pi i r \tau - 2\pi i (\tilde{u}_1\tau -\tilde{u}_2)})(1-e^{2\pi i r \tau + 2\pi i (\tilde{u}_1\tau -\tilde{u}_2)})}\right|^2,
\end{align}
where we denoted $\tilde{u}_1 = u_1 + \frac{a}{N}$ and $\tilde{u}_2 = u_2 + \frac{b}{N}$.
Firstly we shift the integration variables as $u_1 \to u_1 - \frac{a}{N}$, $u_2 \to u_2 - \frac{b}{N}$.
Now we split the holonomy integrals in an integer and fractional part
\begin{equation}
\int_{-\infty}^{+\infty}du_i \to \int_{0}^{1}ds_i \sum_{n_i=-\infty}^{+\infty}
\end{equation}
and we combine the summation over $n_i$ and $a$ (or $b$) in a single sum:
\begin{equation}
\sum_{a=0}^{N-1} \sum_{n_1=-\infty}^{+\infty} \to \sum_{w/N, w=-\infty}^{+\infty}, \quad \sum_{b=0}^{N-1} \sum_{n_2=-\infty}^{+\infty} \to \sum_{m/N, m=-\infty}^{+\infty}.
\end{equation}
In all, we arrive at
\begin{align}
Z &= \frac{1}{N}2\sqrt{k(k-2)}\int_{F}\frac{d\tau d\bar{\tau}}{\tau_2} \int_{0}^{1}ds_1ds_2 \nonumber \\
&\sum_{m,w=-\infty}^{+\infty}\sum_i q^{h_i}\bar{q}^{\bar{h}_i}e^{4\pi\tau_2(1-\frac{1}{4(k-2)}) -\frac{k\pi}{\tau_2}\left|(s_1 + \frac{w}{N})\tau -(s_2 + \frac{m}{N})\right|^2+2\pi\tau_2s_1^2} \nonumber \\
&\frac{1}{\left|\sin(\pi(s_1\tau - s_2))\right|^2}\left|\prod_{r=1}^{+\infty}\frac{(1-e^{2\pi i r \tau})^2}{(1-e^{2\pi i r \tau - 2\pi i (s_1\tau -s_2)})(1-e^{2\pi i r \tau + 2\pi i (s_1\tau -s_2)})}\right|^2.
\end{align}
The next step is the Poisson resummation to extract the discrete momentum quantum number:
\begin{equation}
\sum_{m=-\infty}^{+\infty}e^{-\frac{k\pi}{\tau_2}\left[\frac{m^2}{N^2}-2\frac{m}{N}\left(\left(s_1 + \frac{w}{N}\right)\tau_1 -s_2\right)\right]} = N \sqrt{\tau_2}{k}\sum_{n=-\infty}^{+\infty}e^{-\frac{\pi\tau_2}{k}\left[Nn + \frac{ik}{\tau_2}\left(\left(s_1+\frac{w}{N}\right)\tau_1 - s_2\right)\right]^2}.
\end{equation}
In all, the net effect in comparison to the $N=1$ result is simply
\begin{equation}
n \to N n , \quad w \to w/N.
\end{equation}
This leads to the constraints
\begin{align}
&q-\bar{q} = Nn, \\
&q + \bar{q} + 2j = -k\frac{w}{N}.
\end{align}
The first constraint leads to the conclusion that only discrete momenta $\in N \mathbb{N}$ are allowed. The second constraint combined with a contour-shift argument as given in \cite{Hanany:2002ev}, gives the same unitarity constraints for all sectors as the $N=1$ case:
\begin{equation}
\frac{1}{2} < j < \frac{k-1}{2}.
\end{equation}
Changing the conventions of the $SL(2,\mathbb{R})$ quantum numbers to those we used in section \ref{unitarity}, this condition is the same as (\ref{unibos}).
The remainder of the analysis is the same as the $N=1$ case.
We conclude that the only effect of orbifolding the cigar CFT is including all fractional winding numbers and constraining the discrete momentum. No extra sectors or special cases occur in contrast to the Lorentzian $AdS_3/\mathbb{Z}_N$ case \cite{Son:2001qm}.

\section{Quantum numbers for the bosonic Rindler string}
\label{boswind}
In this appendix we briefly describe the modifications for the bosonic string. For an early treatment of the $SL(2,\mathbb{R})/U(1)$ spectrum, we refer to \cite{Jatkar:1992np}. Here we focus again on the transition from the cigar to flat Euclidean Rindler space. The eigenvalue equation is given by:\footnote{We set $\alpha'=2$ in this section.}
\begin{align}
&-\frac{\partial_\rho\left(\sinh\left(\sqrt{2/(k-2)}\rho\right)\partial_{\rho}T(\rho)\right)}{\sinh\left(\sqrt{2/(k-2)}\rho\right)} \nonumber \\
&\quad \quad \quad  + \left(-2 + w^2\frac{k^2}{2(k-2)}\left(\tanh^2\left(\rho/\sqrt{2(k-2)}\right)-\frac{2}{k}\right)\right)T(\rho)= \lambda T(\rho).
\end{align}
The solution that does not blow up as $\rho \to \infty$ is given by
\begin{equation}
T(\rho) \propto \frac{1}{\cosh\left(\frac{\rho}{L}\right)^{1+\sqrt{\omega}}}\,\, {\mbox{$_2$F$_1$}\left(\frac{\sqrt{\omega}+1+kw}{2},\frac{\sqrt{\omega}+1-kw}{2};\,\sqrt{\omega}+1;\,\frac{1}{\cosh\left(\frac{\rho}{L}\right)^{2}}\right)}.
\end{equation}
where $L = \sqrt{2(k-2)}$ and $\omega = 1-4(k-2)-2(k-2)\lambda +k(k-2)w^2$.
The bound states are again determined when the hypergeometric function reduces to a polynomial. This entails
\begin{equation}
\sqrt{\omega} = kw -2l+1.
\end{equation}
This is again the same condition as the identification with the asymptotic linear dilaton primaries. As an example, the lowest state has $l=1$ and
\begin{equation}
T(\rho) \propto \frac{1}{\cosh\left(\frac{\rho}{\sqrt{2(k-2)}}\right)^{kw}}.
\end{equation}
For large $k$ this gives us again the behavior $e^{-\rho^2/4}$. The eigenvalues are given by
\begin{equation}
\lambda = \frac{-2l(l-1) +2wlk -kw -kw^2 -2k + 4}{k-2},
\end{equation}
whose large $k$ limit gives (upon setting $l=n+1$)
\begin{align}
\lambda &\approx 2wl -w -w^2 -2 \nonumber \\
&= 2wn +w -w^2 -2,
\end{align}
which is indeed the correct result (\ref{bosonicspectrum}). We conclude that also for the bosonic string, $l=n+1$, where $n$ denotes again the quantum number labeling Rindler eigenmodes as in section \ref{bosonic}.

\section{Type II Euclidean Rindler spectrum}
\label{spectr}
In this appendix we discuss the discrete momentum modes and mixed modes (containing both momentum and winding) for type II superstrings in Euclidean Rindler space. We then collect the results in the resulting string spectrum. Our main goal is to identify the cigar quantum numbers with the Rindler quantum numbers. We set $\alpha'=2$ again to lighten the notation.
\subsection*{Discrete momentum states}
The discrete momentum states have wavefunctions obeying the following eigenvalue equation
\begin{equation}
-\frac{\partial_\rho\left(\sinh\left(\sqrt{2/k}\rho\right)\partial_{\rho}T(\rho)\right)}{\sinh\left(\sqrt{2/k}\rho\right)} + \left(-1 + n^2\frac{1}{2k}\coth^2\left(\rho/\sqrt{2k}\right)\right)T(\rho)= \lambda T(\rho)
\end{equation}
where $n$ labels the discrete momentum. The resulting eigenfunctions are almost the same as the winding eigenfunctions (\ref{expl}) except for the replacements $\omega = 1- 2k -2k\lambda + n^2$ and $\cosh \to \sinh$. This is crucial since $\frac{1}{\sinh}$ blows up when its argument goes to zero. This immediately implies that there are no discrete states. The continuous states are determined by a critical eigenvalue
\begin{equation}
\lambda^* = \frac{n^2}{2k} + \frac{1}{2k} -1.
\end{equation}
Taking $k\to\infty$ implies $ \lambda^* \to \frac{n^2}{2k} - 1$ and for $\lambda > \lambda^*$ one finds the continuum. The eigenfunctions one finds in the $k\to \infty$ limit are given by
\begin{equation}
\psi(\rho) \propto J_n\left(\sqrt{1+\lambda}\rho\right) 
\end{equation}
and $1+\lambda$ is indeed positive for the continuous states. The unitarity bounds are not present for the continuous states and all $n \in \mathbb{Z}$ are allowed. Note that even though $J_n$ is damped for large $\rho$, the physical probability density equals $\sqrt{\rho}J_n$ and this does not damp. One can rewrite the wavefunction as
\begin{equation}
\psi \propto J_n\left(\sqrt{t+\frac{n^2}{2k}}\rho\right) 
\end{equation}
with $t \in \mathbb{R}^{+}$. Using the asymptotic behavior, one can identify 
\begin{equation}
t = \frac{2s^2}{k}.
\end{equation} 
In the limit $k \to \infty$, we see that $n$ represents the order of the Bessel function. We see that we should consider only $n < \mathcal{O}(k)$ to have a finite order. This implies the wavefunctions simplify to
\begin{equation}
\psi \propto J_n\left(\sqrt{t}\rho\right) 
\end{equation}
where still $t = \frac{2s^2}{k}$. This implies $s^2$ \emph{can} be of order $k$. These states represent propagating states. One can also see these conclusions in the cigar spectrum:
\begin{equation}
-\frac{\alpha'M^2}{4} + \frac{n^2}{4k} + \frac{s^2+1/4}{k} = 1/2.
\end{equation}
The second term disappears in the large $k$ limit but the third term remains. We find
\begin{equation}
M^2 = \frac{2}{\alpha'}\left(\frac{2s^2}{k} - 1\right) = \frac{2}{\alpha'}\left(t - 1\right),
\end{equation}
exactly analogous to the expressions we found before.\\
We have considered here only the CFT primaries and no oscillators are present here.

\subsection*{Winding and momentum states}
If we relax the condition $L_0^{Rindler} = \bar{L}_0^{Rindler}$, we can find other states. Since we are interested in the one-loop path integral, string states can be off-shell and hence this condition need not be applied. We only require the less strict condition $L_0 - \bar{L}_0 \in \mathbb{Z}$ to preserve modular invariance.\footnote{However, the dominant contribution to the free energy involves also the $\tau_1$ integral. For large $\tau_2$, this integral effectively projects onto states satisfying $L_0 = \bar{L}_0$. Nevertheless, even with this restriction on states, one only requires the total Virasoro operators to satify this property, and a mismatch in the Rindler sector could be compensated by another sector.} The field theory equation of motion is found by writing the Virasoro zero mode as
\begin{align}
L_{0} &= - \frac{1}{k}\left[\partial_r^2 + \coth (r) \partial_r + \frac{1}{4}\coth^2\left(\frac{r}{2}\right)\partial_{\theta}^2 + \frac{1}{4}\tanh^2\left(\frac{r}{2}\right)\partial_{\tilde{\theta}}^2 + \frac{1}{2} \partial_\theta \partial_{\tilde{\theta}}\right], \\
\bar{L}_{0} &= - \frac{1}{k}\left[\partial_r^2 + \coth (r) \partial_r + \frac{1}{4}\coth^2\left(\frac{r}{2}\right)\partial_{\theta}^2 + \frac{1}{4}\tanh^2\left(\frac{r}{2}\right)\partial_{\tilde{\theta}}^2 - \frac{1}{2} \partial_\theta \partial_{\tilde{\theta}}\right].
\end{align}
The third term represents the discrete momentum, the fourth term the winding and the final term mixes these two contributions. If we use a variant of equation (\ref{metricL0}), we get the following effective metric
\begin{equation}
ds^2 = \frac{\alpha' k}{4}\left[dr^2 + \frac{4}{\coth^2\left(\frac{r}{2}\right)}d\theta^2 + \frac{4}{\tanh^2\left(\frac{r}{2}\right)}d\tilde{\theta}^2\right].
\end{equation}
In this metric, coordinates conjugate to the momentum and winding are both present simultaneously. Such a description is very much in the spirit of double field theory \cite{Aldazabal:2013sca}.
The resulting eigenvalue equation one finds is
\begin{align}
&-\frac{\partial_\rho\left(\sinh\left(\sqrt{2/k}\rho\right)\partial_{\rho}T(\rho)\right)}{\sinh\left(\sqrt{2/k}\rho\right)} \nonumber \\
&\quad\quad\quad+ \left[-1 + n^2\frac{1}{2k}\coth^2\left(\rho/\sqrt{2k}\right) + w^2\frac{k}{2}\tanh^2\left(\rho/\sqrt{2k}\right) \right]T(\rho)= \lambda T(\rho),
\end{align}
whose well-behaved solutions are products of $\cosh$, $\sinh$ and hypergeometric functions $\mbox{$_2$F$_1$}$. Again requiring the hypergeometric functions to reduce to polynomials identifies the discrete states by the condition\footnote{We focus on $n \geq 0$.}
\begin{equation}
\sqrt{1-2k-2k\lambda+k^2w^2+n^2} = 1 - kw + n + 2q
\end{equation}
where $q=0,1,2,...$. For instance, when $q=0$, the eigenfunction is given by
\begin{equation}
T(\rho) \propto \frac{\sinh(\rho/\sqrt{2k})^n}{\cosh(\rho/\sqrt{2k})^{kw}},
\end{equation}
whereas for $q=1$ we find
\begin{equation}
T(\rho) \propto \frac{\sinh(\rho/\sqrt{2k})^{n+2}}{\cosh(\rho/\sqrt{2k})^{kw}}\left(1-\coth(\rho/\sqrt{2k})^2\frac{(n+1)}{kw-1}\right).
\end{equation}
The asymptotic behavior of these functions is given by
\begin{equation}
\psi \propto \exp\left(-\frac{2}{\sqrt{k}}\left(\frac{kw}{2} - \frac{n}{2} - q\right)\right),
\end{equation}
and this identifies $q=l+1$. Taking $k$ large\footnote{Either in the differential equation or in the cigar eigenfunctions.}, one finds the corresponding Euclidean Rindler states with
\begin{equation}
T(\rho) \propto \frac{\text{WhittakerM}\left(\frac{n}{2}+\frac{1}{2} + q, \frac{n}{2},w\frac{\rho^2}{2}\right)}{\rho} \propto e^{-w\rho^2/4}\rho^{n+1}L_{q}^{(n)}(w\rho^2/2)
\end{equation}
where $q=0,1,2,...$ and the generalized Laguerre polynomials appear again. These states are localized close to the origin. The eigenvalues are given by
\begin{equation}
\label{eigspectr}
\lambda = wn + 2qw + w -1.
\end{equation}
These states indeed coincide with the large $k$ string spectrum. We should be a little careful in this case. The on-shell condition is
\begin{equation}
L_0 + \bar{L}_0 = 1,
\end{equation}
which reduces to
\begin{equation}
-\frac{\alpha'M^2}{2} + \frac{\left(\frac{kw}{2}+\frac{n}{2}\right)^2}{k} + \frac{\left(\frac{kw}{2}-\frac{n}{2}\right)^2}{k} - 2\frac{\left(\frac{kw}{2}-\frac{n}{2}-l\right)\left(\frac{kw}{2}-\frac{n}{2}-l+1\right)}{k} = 1.
\end{equation}
Taking $k$ large yields
\begin{equation}
-\frac{\alpha'M^2}{2} + \frac{kw^2}{2}  - \frac{kw^2}{2} + wn + 2wl - w = 1
\end{equation}
or
\begin{equation}
M^2 = \frac{2}{\alpha'}\left(wn + 2wl - w - 1\right).
\end{equation}
Setting $l=q+1$, this gives the same spectrum as in (\ref{eigspectr}). To satisfy the unitarity constraints we should again take $w=1$. The quantum numbers are also constrained as $2l > -n+1$, which is trivial. To deal with the $n<0$ states, the reader can readily check that we should simply replace $n \to -n$, such that all formulas work for any $n$ if we would write $n \to \left|n\right|$. \\
Finally the case $w=-1$ has $j = M -l$ with $M = \left|-\frac{k}{2} + \frac{\left|n\right|}{2}\right| = \frac{k}{2} - \frac{\left|n\right|}{2}$.\footnote{Actually, $M$ can be written explicitly as $M = \frac{k\left|w\right|}{2} - \frac{\left|n\right|}{2}$. This expression and the winding-dominance inequality $\left|kw\right| > \left|n\right|$ together are the same restrictions as those displayed in \cite{Dijkgraaf:1991ba}.} The Euclidean Rindler wavefunctions are the same as those of the $w=+1$ case. \\
To sum up, the primaries (in the NS-NS sector) in the type II string spectrum consist of:
\begin{itemize}
\item{Discrete states with $w = \pm 1$, $j= \frac{k}{2}-l$ and $l=1,2,3,\hdots$. 
These states are localized to the origin and have wavefunctions
\begin{equation}
\psi \propto L_{l-1}\left(\rho^2/2\right)e^{-\rho^2/4}.
\end{equation}}
\item{Continuous states with $n \in \mathbb{Z}$ and $j=-1/2 + is$ with $s \in \mathbb{R}$. The wavefunctions are of the propagating form:
\begin{equation}
\psi \propto J_n\left(\sqrt{2/k}\left|s\right|\rho\right). 
\end{equation}}
\item{Discrete states with $w=\pm1$, and $n \neq 0$. Such states are only allowed on-shell if the physical constraint $L_0 - \bar{L}_0 = 0$ can be satisfied by compensating the Rindler quantum numbers by suitable spectator quantum numbers (such as a flat toroidal dimension). 
These states are localized to the origin and are of the form
\begin{equation}
\psi \propto e^{-\rho^2/4}\rho^{\left|n\right|+1}L_{l-1}^{(\left|n\right|)}(\rho^2/2).
\end{equation}
} 
\end{itemize}
All of these states satisfy the GSO projection inherited from the cigar as discussed in \cite{Giveon:2013ica}.

\subsection*{Higher oscillator modes}
Higher oscillator modes are given by the same equation of motion with a shifted mass. This can be seen as follows. All conformal secondaries (these are the oscillator modes) can be obtained by applying the raising operators of the affine Lie algebra to the primary states, e.g.
\begin{equation}
J_{-p}^{b}\left|\psi\right\rangle
\end{equation}
where $p$ is a positive integer and $b$ is a group index. Since we have
\begin{equation}
\left[L_0 , J_{-p}^{b}\right] = p J_{-p}^{b},
\end{equation}
the conformal weights of such states is simply shifted by an amount of $p$ with respect to the primaries. This implies the same equation of motion with a shifted mass term. In particular, all winding modes are localized to the Rindler origin. They are however not massless anymore and so are not really relevant for low energy physics. The equation of motion for such (superstring) oscillator states is effectively given by
\begin{equation}
\left(\tilde{L}_0 + \tilde{\bar{L}}_0 + p - 1\right) \left|\phi\right\rangle = 0,
\end{equation}
where we denoted by $\tilde{L}_0 + \tilde{\bar{L}}_0$ the operator used before that can be written in terms of the Laplacian on the group manifold.

\subsection*{Scaling of thermodynamical quantities of the Euclidean Rindler states}
\begin{figure}[h]
\centering
\includegraphics[width=5cm]{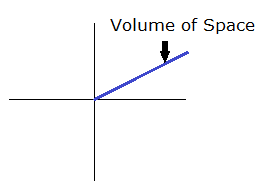}
\caption{The volume of space in Euclidean Rindler space is a half-line starting at the origin: this is the fixed time submanifold.}
\label{volumefig}
\end{figure}

We now ask the question whether the modes we found give contributions to thermodynamical quantities that scale as the volume of the space or as the area transverse to the $\rho$-direction. We previously found that the $w=\pm1$ modes yield a free energy that scales as the transverse area. What about (pure) discrete momentum modes? For this it is convenient to reconsider the heat kernel we explicitly constructed in appendix \ref{appB}. The particle action is given by
\begin{equation}
S = \int_{0}^{T}dt\left(\frac{1}{2}m\dot{\rho}^2 + \frac{1}{2}m\omega^2\rho^2 + \frac{g}{\rho^2}\right),
\end{equation}
with the resulting heat kernel:
\begin{equation}
K(\rho,T|\rho,0) = \rho\left[\frac{m\omega}{\sinh(\omega T)}\right]\exp\left(-m\omega \rho^2\coth(\omega T)\right)I_{\kappa}\left(\frac{\rho^2m\omega}{\sinh(\omega T)}\right)
\end{equation}
The momentum modes are characterized by a positive $g$ and $\omega = 0$. Taking $\omega \to 0$ in the heat kernel yields
\begin{equation}
K(\rho,T|\rho,0) = \rho\left[\frac{m}{T}\right]\exp\left(-\frac{m \rho^2}{T}\right)I_{\kappa}\left(\frac{\rho^2m}{T}\right)
\end{equation}
where $\kappa$ is a strictly positive real number for the momentum modes. This traced heat kernel is sensitive to the size of the space (unlike for the winding modes). This can be seen by evaluating
\begin{align}
\int_{0}^{R} d\rho K(\rho,T|\rho,0) &= \int_{0}^{R} d\rho \rho\left[\frac{m}{T}\right]\exp\left(-\frac{m \rho^2}{T}\right)I_{\kappa}\left(\frac{\rho^2m}{T}\right)
 &= \int_{0}^{\frac{R^2m}{T}} \frac{du}{2} \exp\left(-u\right)I_{\kappa}\left(u\right)
\end{align}
where we regularized the integral by a cut-off at $\rho = R$. One readily checks that the resulting integral is divergent and scales as $R$. The volume of the space is a half-line in the two-dimensional space with a fixed angle since the angular coordinate is a time coordinate. This is depicted in figure \ref{volumefig}.
So we find that the thermodynamical quantities of the discrete momentum modes scale like the volume of the space instead of only the transverse area (like the winding modes).\footnote{Note that this concerns each discrete momentum mode (labeled by $n\in\mathbb{Z}$) separately: we did \emph{not} consider the scaling behavior when summing all of these.} To summarize, the Euclidean Rindler modes give contributions to $F$ and $S$ that scale like
\begin{itemize}
\item{$w=0$. These modes scale like the volume. This can also be seen from their wavefunctions: these modes oscillate and can reach $\rho \to \infty$. Hence they should be sensitive to the entire volume as the above calculation shows.}
\item{$w=\pm1$. These modes scale like the transverse area and these modes are bound to the Euclidean Rindler origin.}
\item{$\left|w\right|>1$. These modes are absent.}
\item{All secondaries (oscillator modes) scale in the same way as the primaries from which they originated (i.e. in the same Verma module).}
\end{itemize}
When considering a cigar geometry instead of flat Euclidean Rindler space, one also has winding modes that scale like the volume (a continuum of states) \cite{Sugawara:2012ag}. The reason is that, while in Euclidean Rindler space the size of the thermal circle keeps increasing, for a cigar it asymptotes to a finite value. This causes a continuum of eigenstates to appear and these scale as the volume of space. These can be interpreted as the asymptotically flat contribution to thermodynamics: the free energy is a sum of two parts, an area-scaling part that gives the contribution of bound states, and a volume-scaling part that gives the usual contribution of a flat space continuum of states \cite{Sugawara:2012ag}.

\end{document}